\documentclass[preprint]{aastex}
\usepackage{lscape,natbib}
\usepackage{amssymb}
\newcommand{\amm}{NH$_3$}
\newcommand{\dia}{N$_2$H$^+$}

\newcommand{\ccs}{C$_2$S}
\newcommand{\hcn}{HC$_5$N}
\newcommand{\kms}{km\,s$^{-1}$}
\newcommand{\cc}{cm$^{-3}$}

\shorttitle{Dense Filaments in Ophiuchus}
\shortauthors{Friesen et al.}

\begin{document}

\title{The Initial Conditions of Clustered Star Formation I: \\ \amm\, Observations of Dense Cores in Ophiuchus}
\author{R. K. Friesen\altaffilmark{1,2}}
\altaffiltext{1}{Department of Physics and Astronomy, University of Victoria, PO Box 3055, STN CSC, Victoria BC CANADA V8W 3P6}
\email{rachel.friesen@nrc-cnrc.gc.ca}

\author{J. Di Francesco\altaffilmark{2,1}}
\altaffiltext{2}{National Research Council Canada, Herzberg Institute of Astrophysics, 5071 West Saanich Road, Victoria, British Columbia, Canada V9E 2E7}

\author{Y. L. Shirley\altaffilmark{3}}
\altaffiltext{3}{Steward Observatory, University of Arizona, 933 N. Cherry Ave., Tucson, AZ 85721}

\and

\author{P. C. Myers\altaffilmark{4}}
\altaffiltext{4}{Harvard-Smithsonian Center for Astrophysics, 60 Garden Street, Cambridge, MA 02138}

\begin{abstract}

We present combined interferometer and single dish telescope data of \amm\, $(J,K)$ = (1,1) and (2,2) emission towards the clustered star forming Ophiuchus B, C and F Cores at high spatial resolution ($\sim 1200$\,AU) using the Australia Telescope Compact Array, the Very Large Array, and the Green Bank Telescope. While the large scale features of the  \amm\, (1,1) integrated intensity appear similar to 850\,\micron\, continuum emission maps of the Cores, on 15\arcsec\, (1800\,AU) scales we find significant discrepancies between the dense gas tracers in Oph B, but good correspondence in Oph C and F. Using the {\sc clumpfind} structure identifying algorithm, we identify 15 \amm\, clumps in Oph B, and 3 each in Oph C and F. Only five of the Oph B \amm\, clumps are coincident within 30\arcsec\, (3600\,AU) of a submillimeter clump. We find $v_{LSR}$ varies little across any of the Cores, and additionally varies by only $\sim 1.5$\,\kms\, between them. The observed \amm\, line widths within the Oph B and F Cores are generally large and often mildly supersonic, while Oph C is characterized by narrow line widths which decrease to nearly thermal values. We find several regions of localized narrow line emission ($\Delta v \lesssim 0.4$\,\kms), some of which are associated with \amm\, clumps. We derive the kinetic temperatures of the gas, and find they are remarkably constant across Oph B and F, with a warmer mean value ($T_K = 15$\,K) than typically found in isolated regions and consistent with previous results in clustered regions. Oph C, however, has a mean $T_K = 12$\,K, decreasing to a minimum $T_K = 9.4$\,K towards the submillimeter continuum peak, similar to previous studies of isolated starless cores. There is no significant difference in temperature towards protostars embedded in the Cores. \amm\, column densities, $N(\mbox{\amm})$, and abundances, $X(\mbox{\amm})$, are similar to previous work in other nearby molecular clouds. We find evidence for a decrease in $X(\mbox{\amm})$ with increasing $N(\mbox{H$_2$})$ in Oph B2 and C, suggesting the \amm\, emission may not be tracing well the densest core gas. 

\end{abstract}


\section{Introduction}

Stars form out of the gravitational collapse of centrally condensed clumps\footnote{In this paper, we call prestellar objects `clumps' instead of `cores' to avoid confusion with the Ophiuchus `Cores' discussed here.} of dense molecular gas. Recent years have seen leaps forward in our understanding of the structure and evolution of isolated, star forming clumps. Most star formation, however, occurs in clustered environments \citep{lada03}. These regions are more complex, with complicated observed geometries, and contain clumps which tend to have higher densities and more compact sizes than those found in isolation \citep{motte98,wardthompson07}. It is likely that due to these differences the star formation process in clustered regions proceeds differently than in the isolated cases. Characterizing the physical and chemical structures of these more complicated regions are thus the first steps towards a better understanding of the process of clustered star formation. 

It is now clear that molecular clumps become extremely chemically differentiated, as many molecules commonly used for tracing molecular gas, such as CO, become severely depleted in the innermost regions through adsorption onto dust grains [see, e.g., \citet{difran07} for a review].  An excellent probe, therefore, of dense clump interiors is the ammonia molecule (\amm), with a relatively high critical density ($n_{cr} \sim 10^4$\,\cc\, for the (1,1) and (2,2) inversion transitions) and apparent resistance to depletion until extreme densities and low temperatures are reached in a starless core's evolution \citep{tafalla04,aikawa05,flower06}. The additional kinematic information provided by line observations are complementary to continuum observations of emission from cold dust, and the ammonia molecule in particular allows the determination of the gas kinetic temperature and density structure due to hyperfine transitions of its metastable states \citep{hotownes}. 

The nearby Ophiuchus molecular cloud, containing the dark L1688 region, is our closest example of ongoing, clustered star formation. The central Ophiuchus cloud has been surveyed extensively in millimeter \citep{young06,stanke06,motte98} and submillimeter \citep{johnstone04,johnstone00} continuum emission. These observations have revealed a highly fragmented complex of star forming clumps with masses $M \sim 0.2 - 6\, M_\odot$, the majority of which are embedded within larger, highly fragmented structures, called `Cores' for historical reasons \citep{loren90} and named A through I, which reside only in areas of high extinction \citep{johnstone04,young06,enoch07}. The total mass of the distinct (sub)millimeter clumps, $\sim 40-50\,M_\odot$, makes up only $0.5 - 3$\,\% of the total $\sim 2000\,M_\odot$ mass of the molecular cloud \citep{young06,johnstone04}.



 Most recent estimates put the distance to the central L1688 cloud region (often also called `$\rho$ Oph') at 120\,pc \citep{loinard08,lombardi08,knude98}, in agreement with some older results \citep{degeus89}, but a clear consensus has not yet been reached. \citet{mamajek08}, for example, find a distance of 135\,pc towards the cloud using Hipparchos parallax data, while VLBA observations by \citet{loinard08} suggest that the Ophiuchus B core may be further distant than the rest of the cloud, at 165\,pc (we also note that Oph B consists of three sub-Cores, B1, B2 and B3, described further in \S 3). This distance is outside the range in median cloud thickness determined by \citet{lombardi08} of $28^{+29}_{-19}$\,pc, but in agreement with an older result by \citet{chini81}. The stars used to determine the distance to Oph B may, however, be background stars \citep{lombardi08}. In the following, we adopt the 120\,pc distance to the entire central Ophiuchus region. 

In this work, we discuss the results of high resolution observations of \amm\, (1,1) and (2,2) in the Ophiuchus B, C and F Cores to study the distribution, kinematics and abundance patterns of the Cores and associated embedded clumps. We find that although the Cores are embedded in the same physical environment, they present very different physical characteristics. We discuss the observations and the combination of interferometer and single dish data in \S 2. In \S 3, we present the data, and detail the hyperfine line fitting procedure and derivations of kinetic temperature $T_K$, \amm\, column $N(\mbox{\amm})$ and space density $n(\mbox{H$_2$})$ in \S 4 (see also Appendix A). We discuss the results in \S 5 and summarize our findings in \S 6. 

\section{Observations and Data Reduction}

\begin{figure}
\plotone{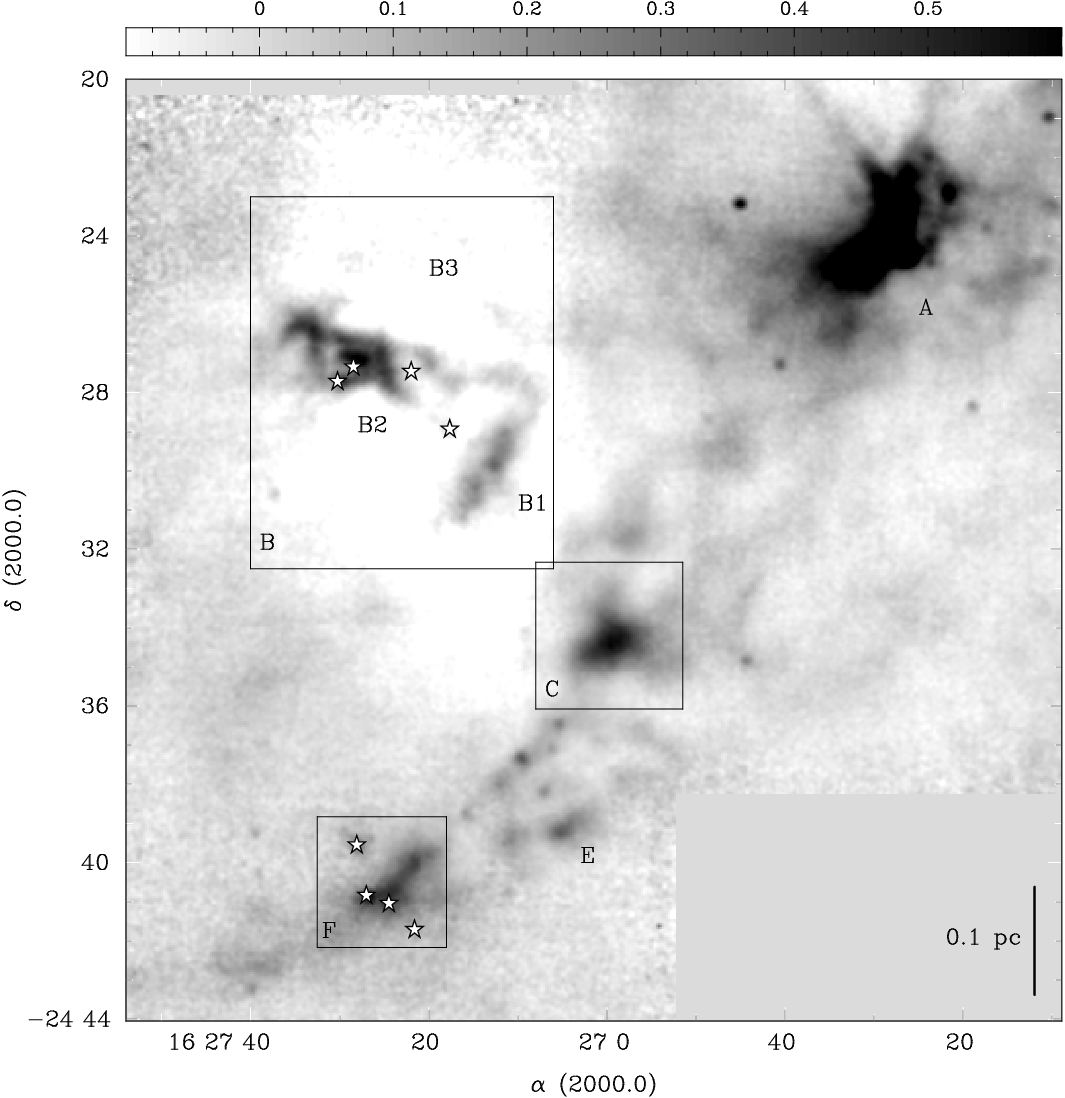} 
\caption{The central region of the Ophiuchus molecular cloud in 850\,\micron\, continuum emission originally mapped at the JCMT by \citet{johnstone00}. Colour scale units are Jy\,beam$^{-1}$, where the beam FWHM $\approx 15$\arcsec. The B, C and F Cores are labelled (Oph B3 is not well detected in 850\,\micron\, continuum). Rectangles show areas mapped in \amm\, (1,1) and (2,2) emission, as well as C$_2$S ($2_1 - 1_0$) and HC$_5$N (9 - 8) at the GBT. Stars indicate locations of protostars identified in the infrared with Spitzer \citep{enoch08}. VLA and ATCA pointings were placed to provide Nyquist-sampled mosaicing of the indicated regions, with multiple beam overlap in the areas of bright continuum emission.}
\label{fig:oph-850}
\end{figure}

Figure \ref{fig:oph-850} shows the central Ophiuchus region in 850\,\micron\, continuum emission first mapped with the Submillimetre Common User Bolometer Array (SCUBA) at the James Clerk Maxwell Telescope (JCMT) by \citet{johnstone00} and recently re-reduced and combined with all other SCUBA archive data in the region by \citet{jorgensen08}, following the description in \citet{kirk06}. The Oph B, C and F Cores are labelled, and boxes show the approximate areas we mapped at the Green Bank Telescope (GBT), the Australia Telescope Compact Array (ATCA) and the Very Large Array (VLA). The details of all astronomical observations are described below. Table \ref{tab:restfreqs} lists the lines observed and their rest frequencies. 

\subsection{Green Bank Telescope}

Single-dish observations of emission from the \amm\ $(J,K)$ = (1,1) and (2,2) inversion lines,  \ccs\, $J_N = 2_1 - 1_0$ and HC$_5$N $J=9-8$ in the Ophiuchus Cores were obtained using the 100\,m Robert C. Byrd Green Bank Telescope (GBT), located near Green Bank, WV, USA.  The observations were done in frequency-switching mode, using the GBT K-band (upper) receiver as the front end, and the GBT spectrometer as the back end. This setup allowed the simultaneous observation of all lines in four 50\,MHz-wide IFs, each with 8192 spectral channels, giving a frequency resolution of 6.104\,kHz, or 0.077\,km\,s$^{-1}$ at 23.694\,GHz. 

The data were taken using the GBT's On-The-Fly (OTF) mapping mode, using in-band frequency switching with a throw of 4\,MHz. In OTF mode, a map is created by having the telescope scan across the target in Right Ascension (R.A.) at a fixed Declination (Dec.), or in Dec. at a fixed R.A., writing data at a predetermined integration interval. The maps of Oph B1 and B2 were made while scanning only in R.A. at a fixed Dec., while for subsequent targets (Oph B3, C and F) the scanning mode was alternated to avoid artificial striping in the final data cubes. No striping, however, is apparent in the final B1 or B2 images. At the observing frequency of 23\,GHz, the telescope beam was approximately 32\arcsec\, FWHM. Subsequent rows or columns were spaced by 13\arcsec\, in Dec. or R.A. to ensure Nyquist sampling. Scan times were determined to ensure either one or two full maps of the observed region could be made between pointing observations. For all observations, pointing updates were performed on the point source calibrator 1622-254 every 45 - 60 minutes, with corrections approximately $2-3$\arcsec. The average telescope aperture efficiency $\eta_A$ and main beam efficiency $\eta_{mb}$ were $0.59 \pm 0.05$ and $0.78 \pm 0.06$ respectively, determined through observations of 3C286 at the start of each shift. The absolute flux accuracy is thus $\sim 8$\,\%. The average elevation of Ophiuchus for all observations was approximately 26 degrees. 

System temperatures ($T_{sys}$) varied between 48\,K and 92\,K over the observation dates with an average $T_{sys} \sim 62$\,K. Table \ref{tab:gbtobsII} gives the area mapped in each region and the final rms sensitivity in K ($T_{MB}$). 

Initial data reduction and calibration were done using the GBTIDL\footnote{GBTIDL is an interactive package for reduction and analysis of spectral line data taken with the GBT.} package. Zenith opacity values for each night were obtained using a local weather model, and the measured main beam efficiency was used to convert the data to units of main beam temperature,  $T_{MB}$. The two parts of the in-band frequency switched data were aligned and averaged, weighted by the inverse square of their individual $T_{sys}$. The data were then converted to AIPS\footnote{The NRAO Astronomical Image Processing System} SDFITS format using the GBT local utility idlToSdfits. In AIPS, the data were combined and gridded using the DBCON and SDGRD procedures. Finally, the data cubes were written to FITS files using FITTP. 

\subsection{Australia Telescope Compact Array}

Maps of \amm\, (1,1) and (2,2) inversion line emission of the Ophiuchus B1, B2, C and F Cores were made over two separate observing runs at the Australia Telescope Compact Array (ATCA). The ATCA is located near the town of Narrabri in New South Wales, Australia, and consists of six antennas, each 22\,m in diameter. Five of the six antennas are movable along the facility's east-west line and small north line, while the sixth antenna is permanently placed along the east-west line at a distance of 6\,km from the other antennas.  An 8\,MHz bandwidth with 1024 channels was used, which provided a spectral resolution of 7.81\,kHz (0.1\,\kms\, at 23.694\,GHz). This configuration enabled in-band frequency switching for the observations. The Oph B Cores were observed over four 9-hour tracks of the array (August 5 - 8, 2004), and the Oph C and F Cores were observed over three 9-hour tracks of the array (May 5 - 7, 2005). Both sets of observations were done with the array in its H168 configuration. This is a compact, hybrid configuration, where three of the movable antennas are placed along the east-west line and two antennas are located on the north spur. Baselines ranged from 61.2\,m to 184.9\,m ($\sim 4.7$\,k$\lambda$ - 14.2\,k$\lambda$) with five antennas. At 23.7\,GHz, these observations provided a primary beam (field-of-view) of 2\arcmin.

Maps were made of the cores using separate pointings spaced by $\Theta_N = ( 2\,/\,\sqrt{3} ) \, \lambda \, / \, 2D \sim 1.1\arcmin$\, at 23.7\,GHz for Nyquist sampling on a hexagonal grid. Table \ref{tab:intobs} gives the number of individual pointings required to cover each Core, the multiple-beam overlap area observed, and the final rms sensitivity achieved towards each Core for both the ATCA and the VLA observations (described below). 

Observations cycled through each individual pointing between phase calibrator observations to maximize $uv$-coverage and minimize phase errors. The phase calibrator, 1622-297, was observed every 20 minutes, and was also used to check pointing every hour. Flux and bandpass calibration observations were performed every shift on the bright continuum sources 1253-055, 1934-638 and 1921-293. 

The ATCA data were reduced using the MIRIAD data reduction package \citep{sault95}. The data were first flagged to remove target observations unbracketed by phase calibrator measurements, data with poor phase stability or anomalous amplitude measurements. The majority of the data were good, as the weather was stable during the observations. Much of the data from baselines involving the 6\,km antenna, however, were flagged due to poor phase stability. The bandpass, gains and phase calibrations were applied, and the data were then jointly deconvolved.  First, the data were transformed from the spatial frequency ($u,v$) plane into the image plane using the task INVERT and natural weighting to maximize signal to noise. The data were then deconvolved and restored using the tasks MOSSDI and RESTOR to remove the beam response from the image. MOSSDI uses a Steer-Dewdney-Ito CLEAN algorithm \citep{steer84}. The cleaning limit was set at twice the rms noise level in the beam overlap region for each object. Clean boxes were used to avoid cleaning noise in the outer regions with less beam overlap. Applying natural weighting provided a final synthesized beam of $\sim 8\arcsec\, \times 10$\arcsec\, FWHM. 

\subsection{Very Large Array}

Maps of \amm\ (1,1) and (2,2) emission were made at the Very Large Array (VLA) near Socorro, NM, USA over the period of 2007 January 20 through 2007 February 11. Nine observing shifts were allotted to the project in the array's DnC configuration, each five hours in duration covering the LST range 14:00 - 19:00. The DnC configuration is a hybrid of the most compact D configuration with the next most compact C configuration. This setup ensures a more circular beam shape for southern sources like Ophiuchus while retaining the sensitivity of the D configuration. 

For these observations, we used a correlator setup with two IFs, each with a bandwidth of 3.125\,MHz with 24.414\,kHz spectral resolution (0.3\,\kms). While not providing the same high spectral resolution obtained at the ATCA, this setup enabled simultaneous observations of the (1,1) and (2,2) lines, and allowed the main hyperfine component and the two middle satellite components of the (1,1) line to be contained within the band. Mosaic maps were made with Nyquist-spaced ($\sim 1.0$\arcmin\, at 23\,GHz) individual pointings. Observations cycled through pointings between phase calibrator observations. Table \ref{tab:intobs} gives the pointing and rms sensitivity information for each Core. 

During these observations, several antennas in the array had been upgraded as part of the Expanded VLA (EVLA) project. Online Doppler tracking at this time was not yet available, and the observations were thus obtained in a fixed frequency mode, with line sky frequencies calculated using the NRAO's online Dopset tool. The observing frequencies were updated frequently, with phase calibrator observations on 1625-254 before and after a frequency change to avoid phase jump problems. Bandpass and absolute flux calibration were done for each shift using observations of 1331+305. 

The data were checked, flagged and calibrated using the NRAO Astronomical Image Processing System (AIPS), following the procedures outlined in the AIPS Cookbook\footnote{http://www.aips.nrao.edu/CookHTML/CookBook.html}. In addition, special processing was required to account for differences in bandpass shape and antenna sensitivity between the VLA and EVLA antennas. System temperatures were lacking for EVLA antennas, so the VLA back-end $T_{sys}$ values were used when reading the data into an AIPS $uv$-database. A bandpass table was then created from the line dataset and applied to a spectrally averaged `channel 0' dataset. The normal VLA calibration steps were then followed.The calibrated $uv$ files were then written to FITS format and imported to MIRIAD, where the data were deconvolved and restored. Applying natural weighting and a taper of 8\arcsec\, $ \times $ 6\arcsec\, provided a final synthesized beam of $\sim 10.5\arcsec\, \times 8.5$\arcsec\, FWHM. 

\subsection{Combining Single Dish and Interferometer Data Sets}

Since none of the antennas in the ATCA and the VLA can act as single radio dishes, there is necessarily an upper limit to the size of structure to which each is sensitive. This upper scale limit is dependent on the shortest spacing between two antennas in the array, and the missing information is thus referred to as the short- or zero-spacing problem. Mosaicing helps to recover short spacing information. For complex sources with emission on many scales, however, combining interferometer data with single-dish observations provides more complete coverage of the $uv$  plane and thus creates a more accurate representation of the true source emission structure. Ideally, the single dish diameter should be larger than the minimum interferometer baseline to ensure maximal overlap in the $uv$-plane and determine accurately flux calibration factors between the observations.

Data from each interferometer were first combined separately with the single dish observations. The GBT data were regridded to match the interferometer data in pixel scale and spectral resolution. The data were then converted to units of Jy\,beam$^{-1}$ for combination with the interferometer data using the average beam FWHM measured at the GBT during the observations. Combination of the data was done using MIRIAD's IMMERGE task. IMMERGE combines deconvolved interferometer data with single-dish observations by Fourier transforming both datasets and combining them in the Fourier domain, applying tapering functions such that at small spacings (low spatial frequencies) the single-dish data are more highly weighted than the interferometer data, while conversely at high spatial frequencies the interferometer data are weighted more highly. The flux calibration factor between the two datasets was calculated in IMMERGE by specifying the overlapping spatial frequencies between the GBT and the interferometers. We took the overlap region in both cases to be 35\,m - 100\,m (2.7\,k$\lambda$ - 7.7\,k$\lambda$), yielding a flux calibration factor of $\sim 1.4$ between the GBT and ATCA datasets, and $\sim 1.0$ between the GBT and VLA datasets in the \amm\, (1,1) line emission. These factors were then applied to the \amm\, (2,2) datasets. The final resolution of the combined data is the same as that of the interferometer data. 

To combine all three datasets, the ATCA and VLA data were imaged together using the INVERT task, applying natural weighting and taper as described for the VLA imaging. The interferometer data were then cleaned and combined with the GBT data as described above. The overlap region was taken to be 35\,m - 100\,m, yielding a flux calibration factor of $\sim 1.3$. By convolving the combined data to match the 32\arcsec\, resolution of the GBT data, we estimate the total flux of the combined image recovers nearly all ($\sim 98$\,\%) the flux in the single-dish image. These data were used to identify structures in the \amm\, data cubes using an automated structure finding algorithm (see \S 3). The data were also tapered to a slightly lower resolution of 15\arcsec\, FWHM to provide higher signal-to-noise ratios for a multiple component hyperfine line fitting routine. 

\section{Results}

\subsection{Comparison with submillimeter dust continuum emission}

We first discuss the \amm\, (1,1) intensity in the combined datasets, and compare the distribution of \amm\, emission and 850\,\micron\, continuum emission in the Oph Cores as shown in Figure \ref{fig:oph-850}. Figures \ref{fig:b-compare}a, \ref{fig:c-compare}a and \ref{fig:f-compare}a show the combined \amm\, (1,1) line emission at $\sim 10.6\arcsec\, \times 8.5$\arcsec\, FWHM resolution. The emission has been integrated over the central hyperfine components in the Oph B, C and F Cores, respectively, with a clip of $\sim 2 \times$ the map rms noise level. (Since the outer edge of the combined maps have higher rms noise levels, integrating only over, i.e., the `main component' of \amm\, reduced the amount of signal included in the noisy outer regions.) The respective 850\,\micron\, emission for each Core at $\sim 15$\arcsec\, FWHM resolution, i.e., lower than the resolution of the combined \amm\, data, is shown in Figures \ref{fig:b-compare}b, \ref{fig:c-compare}b and \ref{fig:f-compare}b. We also show locations of submillimeter clumps identified with the 2D version of {\sc clumpfind} \citep{jorgensen08}. In Figure \ref{fig:b-compare}b, we additionally label the $\sim 1$\,M$_\odot$ continuum object MM8 \citep{motte98}. 

Figures \ref{fig:b-compare}, \ref{fig:c-compare} and \ref{fig:f-compare} also show locations and labels of `cold' YSOs (based on bolometric temperatures derived from fitting their spectral energy distributions) detected and classified through Spitzer infrared observations \citep{enoch08} overlaid on the submillimeter and \amm\, emission. The objects plotted were all identified as Class I protostars (no Class 0 protostars have been associated with Oph B, C or F). Oph B2 is associated with three YSOs. Of these, two are previously known (IRS45/GY273 and IRS47/GY279). Based on association with a continuum emission object \citep{enoch08} and observed infrared colours \citep{jorgensen08}, either three or two protostars in Oph B2 are embedded in gas and dust. One additional Class I protostar is located between B1 and B2, and is only associated with diffuse \amm\, emission. Oph B1 and B3 appear starless. No embedded protostars are associated with Oph C by \citet{enoch08}, but a ``Candidate YSO" is identified by \citep{jorgensen08} 30\arcsec\, south of the Core continuum peak (R.A. 16:26:59.1, Dec. -24:35:03). Based on the different classifications by the two papers, the significant offset from the continuum emission peak, and the lack of any clear influence on the gas in our data, we will discuss Oph C assuming it is not associated with a deeply embedded object. Four protostars are associated with Oph F, all of which have been previously identified (see Figure \ref{fig:f-compare} for object names). Three are likely embedded in the Core. 

Peaks of \amm\, integrated intensity can be used to surmise the presence of `objects', but such identifications can ignore any differences in velocity between adjacent dense gas. We therefore used the 3D version of the automated structure-finding routine {\sc clumpfind} \citep{williams94} to identify distinct \amm\, emission objects in the combined \amm\, (1,1) data cube, which we will henceforth call ``\amm\, clumps". {\sc clumpfind} uses specified brightness contour intervals to search through the data cube for distinct objects identified by closed contours. The size of the identified clumps are determined by including adjacent pixels down to an intensity threshold, or until the outer edges of two separate clumps meet. The clump location is defined as that of the emission peak. {\sc clumpfind} was used only on the main emission component, where multiple hyperfine components are sufficiently blended to present effectively a single line given the 0.3\,\kms\, spectral resolution and typically wide line widths found. The standard interval of $2 \times$ the data rms between contours, with a slightly larger threshold value worked well to separate distinct emission components in all clumps. A lower limit of 1\,K with intervals of 0.4\,K, $\sim 2\,\times$ the cube rms, identified separate emission peaks sufficiently for Oph B and C. Oph F was sufficiently fit using a lower limit of 1.2\,K and intervals of 0.65\,K, as the rms of the combined data was slightly higher. Even so, some identified clumps appeared by eye to be noise spikes at the map edges, and these were culled from the final list, as well as any clumps which contained fewer pixels than in the synthesized beam. Table \ref{tab:peakdat} lists the locations, FWHMs, effective radii and peak brightness temperatures for  {\sc clumpfind}-identified cores in Oph B, C and F, and the clump locations are overlaid on Figures \ref{fig:b-compare} - \ref{fig:f-compare}. The centroid velocities and line widths of the \amm\, clumps were determined through spectral line fitting (described further in \S4), which provided more accurate measures of $v_{LSR}$ and $\Delta v$ given the hyperfine structure of the \amm\, lines. 

In Oph B, Figure \ref{fig:b-compare} shows that the large-scale structure of the Core is similar in both line and continuum emission, but some significant differences are also apparent. For example, the integrated line emission displays a less pronounced division between the B1 and B2 Cores seen in the continuum, and indeed shows significant filamentary structure in the region connecting B1 and B2 where little to no continuum emission is observed. Similarly, strong line emission is present in the south-eastern edge of B2 where the continuum map shows relatively little emission. Additionally, while B2 is the stronger continuum emitter, B1 is significantly brighter in integrated line emission. Strong \amm\, emission in the northern half of B1 extends beyond the continuum contours, and becomes significantly offset from the bulk of the continuum emission to the east . The \amm\, (1,1) observations of the Oph B core region also reveal a narrow-line emission peak north of the western edge of B2, which is coincident with a DCO$^+$ object, Oph B3 \citep{loren90}. B3 is just visible at the lowest contour in the \amm\, integrated intensity map but is not visible in the continuum map. 

Furthermore, although the peaks of continuum emission and line emission are often located in the same vicinity  in B1 and B2, the brightest continuum peaks and the integrated line emission maxima are typically non-coincident. Overall, the mean separation between an \amm\, integrated intensity peak and the nearest 850\,\micron\, continuum peak in B1 and B2 is $\sim 22$\arcsec\, ($2600$\,AU), or $\sim 2 \times$ the \amm\, FWHM resolution. Fifteen {\sc clumpfind}-identified \amm\, clumps are identified in the combined Oph B map, with an average minimum separation between \amm\, clumps of 47\arcsec\, ($5600$\,AU, or 40\arcsec\, ($4800$\,AU), if Oph B3 is not included). The mean minimum distance between \amm\, clumps and submillimeter continuum clumps is 44\arcsec\, (5300\,AU), or $\sim 4 \times$ the \amm\, FWHM resolution. Only five of the fifteen \amm\, clumps are located within 30\arcsec\, (3600\,AU) of a submillimeter continuum clump.

No protostars in Oph B are found at positions of \amm\, (1,1) integrated intensity maxima nor are they associated with identified \amm\, clumps. One protostar, IRS47/GY279, is located south of the \amm\, clump we identify as B2-A6. The offset between the \amm\, clump peak and the protostar, however, is $\sim 30$\arcsec, or approximately $3 \times$ the angular resolution of the combined \amm\, data. A second protostar, IRS45/GY273, is coincident with a submillimeter continuum peak but has little associated \amm\, emission. Either of these protostars may be the source of an east-west aligned outflow recently proposed by \citet{kamazaki03} from CO observations of B2. The third, previously unidentified protostar, seen west of \amm\, clump B2-A4, is located within a narrow ($\sim 30\arcsec$) \amm\, integrated intensity {\it minimum} between the B2 core and the filament connecting Oph B1 and B2. 

Figure \ref{fig:c-compare} shows that Oph C has more similar overall structure when traced by the continuum and integrated \amm\, (1,1) line emission than Oph B. Extended emission in Oph C is elongated along a southeast-northwest axis and contains a single integrated intensity peak. A thin ($\sim 30$\arcsec) filament of faint emission extends to the north of the central peak. The submillimeter continuum emission is largely coincident with the integrated \amm\, contours, but the continuum emission peak is offset to the \amm\, integrated intensity peak by $\sim 20$\arcsec. {\sc clumpfind} separates the central \amm\, emission into two cores, C-A1 and C-A3, and finds a third object, C-A2, at the tip of the northern emission extension. C-A1 and C-A3 are found to the northwest and southeast (30\arcsec\, offset and 15\arcsec\, offset, respectively) of the centres of both the \amm\, integrated intensity and of the continuum emission.  Continuum emission also extends in the direction of the faint \amm\, northern extension, but there is no secondary peak present. 

Like Oph C (but unlike Oph B), Figure \ref{fig:f-compare} shows that Oph F also has very similar structure when traced by either the submillimeter continuum emission or integrated \amm\, intensity. Unlike both Oph B and C, the {\sc clumpfind}-identified \amm\, clumps are coincident with the integrated intensity peaks. F-A3 is nearly coincident (within a beam FWHM) with a submillimeter continuum clump and is additionally coincident with an embedded protostar, IRS43/GY265. F-A2 is associated with a second embedded protostar (CRBR65) and continuum emission, but not an identified continuum clump. A thin filament ($\sim 15$\arcsec) extends to the northwest and the third \amm\, clump, F-A1, which is also coincident with extended continuum emission but no identified clump. A third embedded protostar (IRS44/GY259) in the north-east is coincident with a continuum peak, but has no associated \amm\, emission. A fourth protostar, in the south-west, may be coincident with some unresolved \amm\, emission, but is located in a section of the map with larger rms values and consequently the small integrated intensity peak seen at that location may be simply noise. 

The discrepancies between \amm\, and submillimeter continuum emission in Oph B are in contrast to earlier findings of extremely high spatial correlation between the two gas tracers in isolated, low-mass starless clumps. For example, \citet{tafalla02} found that both the millimetre continuum and integrated \amm\, (1,1) and (2,2) line intensity were compact  and centrally concentrated in a survey of five starless clumps, including L1544 in the Taurus molecular cloud. In these clumps, the integrated intensity maxima of both transitions are approximately coincident (within the 40\arcsec\, angular resolution of the \amm\, observations) with the continuum emission peaks. This same coincidence between \amm\, and millimeter continuum was found in B68 \citep{lai03}. When observed at higher angular resolution, \amm\, emission in L1544 remained coincident with the continuum but the line integrated intensity peak was offset by $\sim 20$\arcsec. This offset was explained, however, as being due to the the \amm\, emission becoming optically thick \citep{crapsi07}. 

It is also possible that different methods of identifying structure in molecular gas (such as the {\sc gaussclump} method of \citealp{stutzki90}, or using dendrograms as in \citealp{rosolowsky08d}, for example) would create a different `core' list than presented here. We have additionally compared our results in Oph B with the locations of millimeter objects identified using multi-wavelet analysis by \citet{motte98}, and find a smaller yet still significant mean minimum distance of 27\arcsec\, between \amm\, clumps and millimeter objects. Given the severe positional offsets in some locations between the \amm\, emission and submillimeter continuum, it is unlikely that different structure-finding methods would provide substantially different results. 

\subsection{Single Dish C$_2$S and HC$_5$N Detections}

\begin{figure}
\plotone{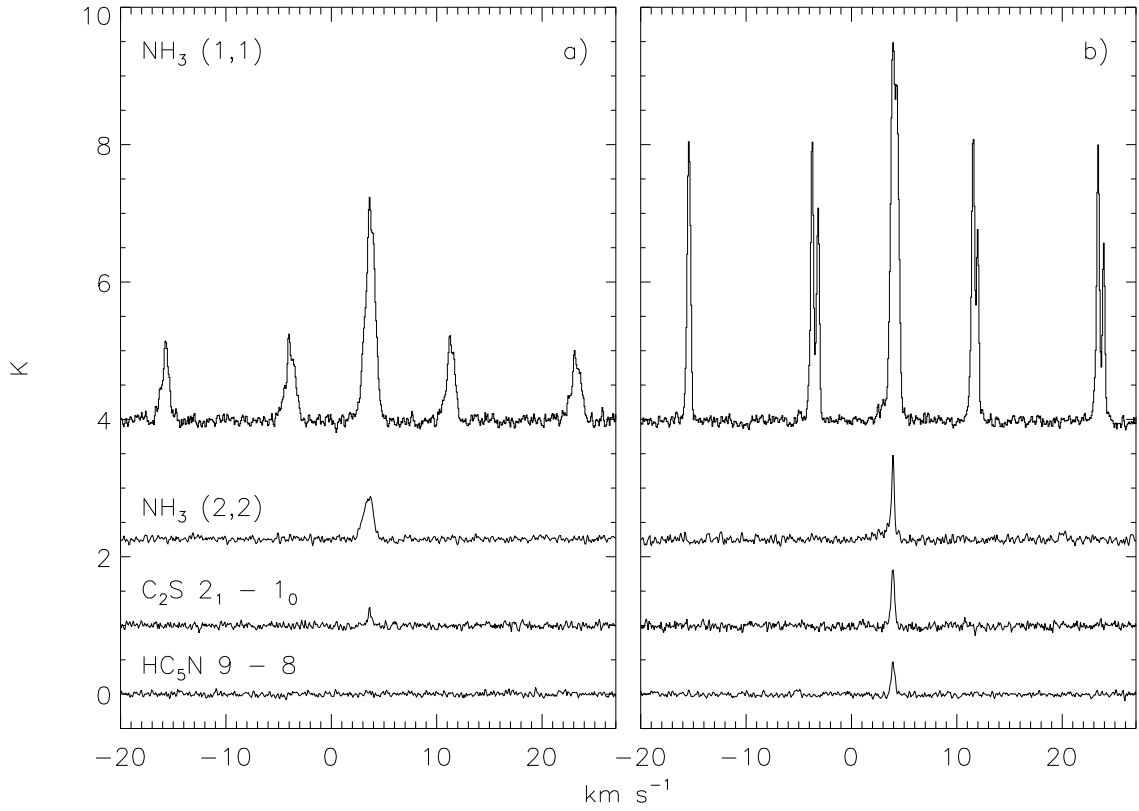} 
\caption{a) Spectra of all species observed at the GBT at the integrated intensity C$_2$S $(2_1 - 1_0)$ peak in Oph B1. The \amm\, (1,1) and (2,2) and C$_2$S baselines are offset from 0 for clarity. b) Spectra of all species observed at the GBT at the integrated intensity C$_2$S $(2_1 - 1_0)$ peak in Oph C. The \amm\, (1,1) and (2,2) and C$_2$S baselines are offset for clarity.}
\label{fig:b1c-spec}
\end{figure}

\begin{figure}
\plotone{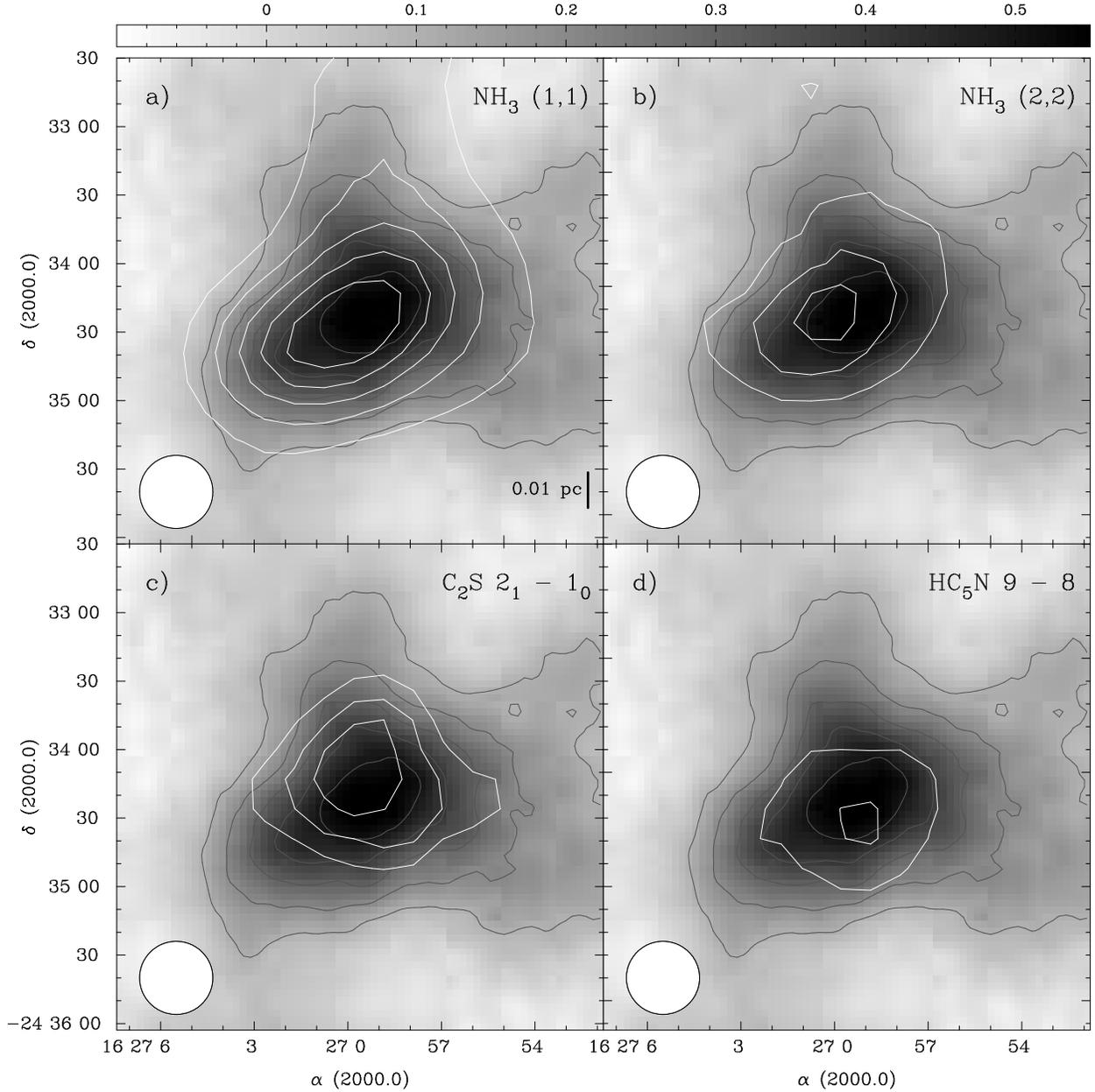} 
\caption{GBT observations of Oph C. The circle in each panel gives the relative FWHM of the GBT data. a) The Ophiuchus C core in 850\,\micron\, continuum emission at 15\arcsec\, resolution (greyscale, Jy\,beam$^{-1}$ units) overlaid with continuum contours (grey). Black contours show the integrated \amm\, (1,1) intensity at $\sim 32$\arcsec\, resolution, beginning at 3\,K\,\kms\, and increasing by 1.5\,K\,\kms. For all plots, emission was integrated over spectral channels with intensities $\geq \pm 2\,\sigma$ rms. b) 850\,\micron\, continuum emission in Oph C overlaid with black \amm\, (2,2) integrated emission contours. Contours begin at 0.3\,K\,\kms\, and increase by 0.1\,K\,\kms. c) 850\,\micron\, continuum emission in Oph C overlaid with black C$_2$S $2_1 - 1_0$ integrated emission contours. Contours are 0.2, 0.25, 0.3\,K\,\kms. d) 850\,\micron\, continuum emission in Oph C overlaid with black HC$_5$N $9-8$ integrated emission. Contours are 0.1, 0.2\,K\,\kms.} 
\label{fig:c-gbt-all}
\end{figure}

The GBT observations of \ccs\, $2_1 - 1_0$ emission only resulted in single, localized detections in Oph B1 and Oph C, while only Oph C had a single, localized detection in \hcn\, (9-8). The \ccs\, emission in B1 was confined to a single peak at its southern tip. The single-dish spectra of all observed species at the \ccs\, peak locations in B1 and C are presented in Figure \ref{fig:b1c-spec} (note that in Figure \ref{fig:b1c-spec}, the \amm\, line is so narrow in Oph C that we are detecting the hyperfine structure of the (1,1) line). The integrated intensity GBT maps of all molecules observed in Oph C are shown in Figure \ref{fig:c-gbt-all}. Within the $\sim 30$\arcsec\, resolution limits of the GBT data, the \amm, \ccs\, and \hcn\, spectral line integrated intensity peaks overlap with the local 850\,\micron\, continuum emission peak in Oph C. We fit the spectra of the two \ccs\, and single \hcn\, detections with single Gaussians to determine their respective $v_{LSR}$, line width $\Delta v$, and peak intensity of the line in $T_{MB}$ units. We additionally fit a 2D Gaussian to the integrated intensity maps to determine the FWHMs of the emitting regions. We find a beam-deconvolved FWHM = 9200\,AU and 5800\,AU for the \ccs\, and \hcn\, emission, respectively, in Oph C. The \ccs\, emission in Oph B1 is elongated, with a FWHM = 9400\,AU in R.A. but only 3700\,AU in Dec. for an effective FWHM = 5100\,AU. The results of the Gaussian fitting are listed in Table \ref{tab:ccsdat}. See \S 4.4.3 for further analysis of \ccs\, and \hcn. 

\section{\amm\, Line Analysis}

\subsection{\amm\, Hyperfine Structure Fitting}
The metastable $J = K$ rotational states of the symmetric-top \amm\, molecule are split into inversion doublets due to the ability of the N-atom to quantum tunnel through the hydrogen atom plane. Quadrupole and nuclear hyperfine effects further split these inversion transitions, resulting in hyperfine structure of the ($J,K$) = (1,1) transition, for example, containing 18 separate components. This hyperfine structure allows the direct determination of the optical depth of the line through the relative peaks of the components. Additionally, since transitions between $K$-ladders are forbidden radiatively, the rotational temperature describing the relative populations of two rotational states, such as the (1,1) and (2,2) transitions, can be used to determine directly the kinetic gas temperature \citep{hotownes}. 

For a given (J,K) inversion transition of \amm\, in local thermodynamic equilibrium (LTE), the observed brightness temperature $T_A^*$ as a function of frequency $\nu$ can be written as 

\begin{equation}
T_{A,\nu}^*(J,K)  = \eta_{MB} \, \Phi \, (J(T_{ex}(J,K)) - J(T_{bg})) (1 - \exp(-\tau_\nu (J,K)))
\label{eqn:tastar}
\end{equation}

\noindent assuming the excitation conditions of all hyperfine components are equal and constant (i.e., $T_{ex,\nu} = T_{ex}$). Here, $\eta_{MB}$ is the main beam efficiency, $\Phi$ is the beam filling factor of the emitting source, $T_{ex}$ is the line excitation temperature, $T_{bg} = 2.73$\,K is the temperature of the cosmic microwave background, and $J(T) = (h\nu/k) [\exp(h\nu/kT) - 1]^{-1}$. The line opacity as a function of frequency, $\tau_\nu$, is given by

\begin{equation}
\tau_\nu  = \tau_0 \sum_{j=1}^{N} a_j \exp\biggl(-4\ln2\,\biggl(\frac{\nu-\nu_0 - \nu_j}{\Delta \nu}\biggr)^2\biggr)
\label{eqn:tau}
\end{equation}

\noindent where $N$ is the total number of hyperfine components of the (J,K) transition ($N = 18$ for the (1,1) transition and $N=21$ for the (2,2) transition). For a given $j^{th}$ hyperfine component, $a_j$ is the emitted line fraction and $\nu_j$ is the expected emission frequency. The observed frequency of the brightest line component is given by $\nu_0$, with a FWHM $\Delta \nu$. Values of $a_j$ and $\nu_j$ were taken from \citet{kuko67}.  Here, we assume $\Phi = 1$. If the observed emission does not entirely fill the beam, the determined $T_{ex}$ will be a lower limit. In regions where the emission is very optically thin ($\tau << 1$) there is a degeneracy between $\tau$ and $T_{ex}$ and solving for the parameters independently becomes impossible. We restricted our analysis to regions where the \amm\, (1,1) intensity in the central component is greater than 2\,K, which corresponds roughly  in our data to a signal-to-noise ratio of $8-10$ in the main component and $\sim 2-3$ in the satellite components. With this restriction, we also find $\tau \gtrsim 0.5$ throughout the regions discussed. 

To improve the signal-to-noise ratio of the data and to match the resolution of the 850\,\micron\, continuum data, we first convolved the combined data to a final FWHM of 15\arcsec\, (from 10.6\arcsec\, $\times$ 8.5\arcsec), and then binned the convolved data to 15\arcsec\, $\times$ 15\arcsec\, pixels. Assuming Gaussian profiles, the 18 components of the \amm\, (1,1) emission line were fit simultaneously using a chi-square reduction routine custom written in {\sc idl}. The returned fits provide estimates of the line centroid velocity ($v_{lsr}$), the observed line FWHM ($\Delta v_{obs}$), the opacity of the line summed over the 18 components ($\tau$), and $[J(T_{ex}) - J(T_{bg})]$. The satellite components of the (2,2) line are not visible above the rms noise of our data. These data were consequently fit (again in {\sc idl}) with a single Gaussian component. 

The line widths determined by the hyperfine structure fitting routine are artificially broadened by the velocity resolution (0.3\,\kms) of the observations. To remove this effect, we subtract in quadrature the resolution width, $\Delta v_{res}$, from the observed line width, $\Delta v_{obs}$, such that $\Delta v_{line} = \sqrt{\Delta v_{obs}^2 - \Delta v_{res}^2}$. In the following, we simply use $\Delta v = \Delta v_{line}$ for clarity. The limitations of the moderately poor velocity resolution are discussed further in Appendix \ref{ap_res}, but do not significantly impact our analysis. In regions where lines are intrinsically narrow, such as Oph C and parts of Oph F, the derived line widths may be overestimated by up to $\sim 20 - 35$\,\%. 

The uncertainties reported in the returned parameters are those determined by the fitting routine, and do not take the calibration uncertainty of $\sim 8$\,\% into account. The calibration uncertainty affects neither the derived parameters that are dependent on ratios of line intensities, such as the opacity and kinetic temperature $T_K$, nor the uncertainties returned for $v_{LSR}$ or $\Delta v$. The excitation temperature, however, as well as the column densities and fractional \amm\, abundances discussed below (see \S 4.4) are dependent on the amplitude of the line emission, and are thus affected by the absolute calibration uncertainty. 

Table \ref{tab:linefit} lists the mean, rms, minimum and maximum values of $v_{lsr}$, $\Delta v$, $\tau$ and $T_{ex}$ found in each of the Cores using the above restrictions for the combined \amm\, line emission. In the following sections, we describe in detail the results of the line fitting and examine the resulting line centroid velocities and widths, as well as $T_{ex}$ and $\tau$. In addition, we use the fit parameters to calculate the gas kinetic temperature ($T_K$), non-thermal line widths ($\sigma_{NT}$), \amm\, column density ($N(\mbox{\amm})$), gas density ($n(\mbox{H$_2$})$) and \amm\, abundance ($X(\mbox{\amm})$) across all the cores, as described further in \S 4.2, 4.3 and 4.4. Table \ref{tab:tketc} summarizes the mean, rms, minimum and maximum values of $T_K$, $\sigma_{NT}$, $N(\mbox{\amm})$, $n(\mbox{H}_2)$ and $X(\mbox{\amm})$ for each Core. For each \amm\, clump, Table \ref{tab:peak_dat} summarizes the mean, rms, minimum and maximum values of all determined parameters (means were obtained by uniformly weighting each pixel). 

\subsection{Line Centroids and Widths}

\begin{figure}
\epsscale{0.75}
\plotone{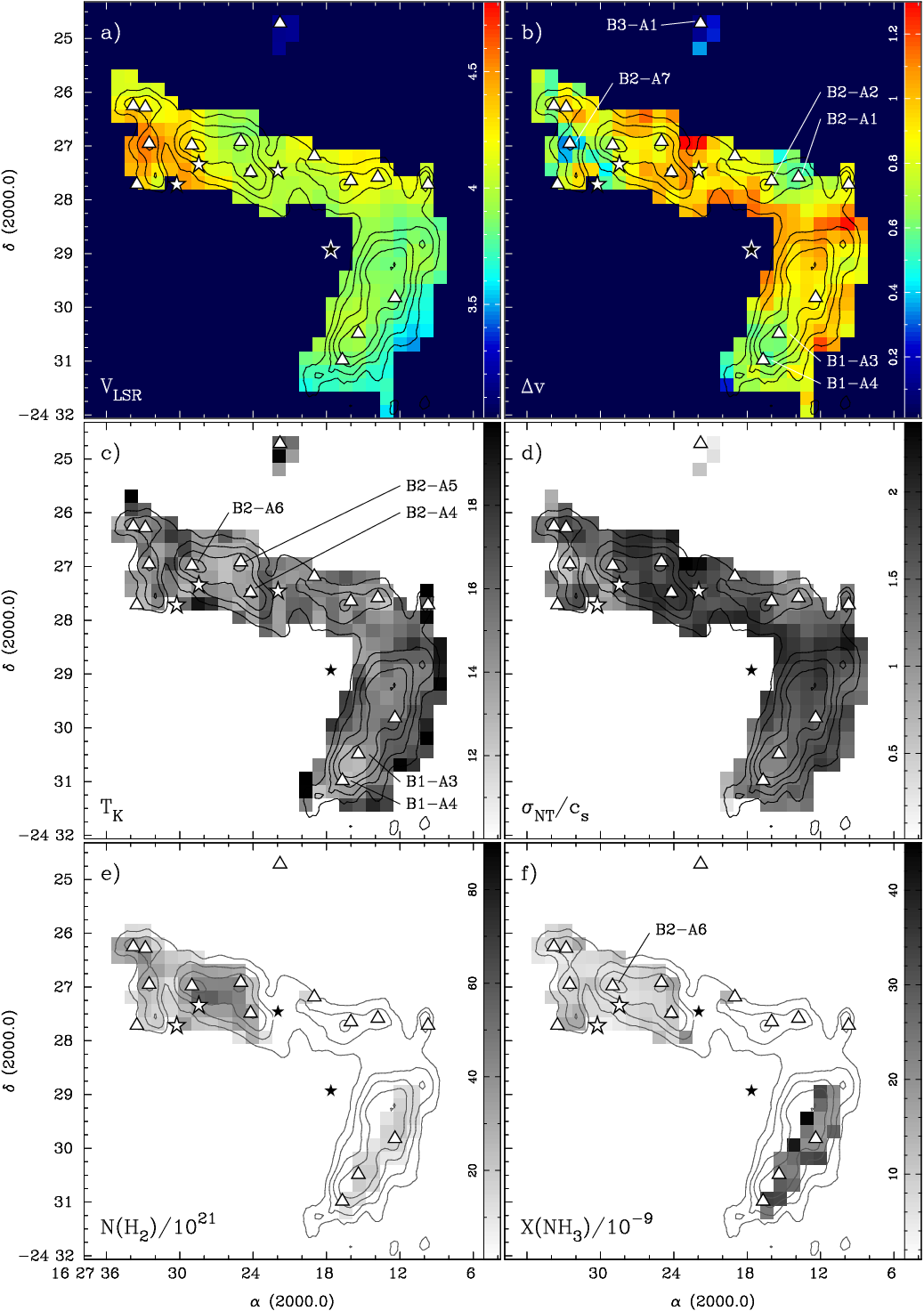}
\caption{a) Line velocity or $v_{LSR}$ in Oph B. Colour scale is in \kms. The values shown are those derived after convolving the data to 15\arcsec\, resolution and then regridding into 15\arcsec\, $\times$ 15\arcsec\, pixels. In all plots, contours show integrated \amm\, (1,1) intensity convolved to 15\arcsec\, resolution, beginning at 4.5\,K\,\kms\, and increasing by 1.5\,K\,\kms. Stars indicate protostar positions and triangles indicate positions of \amm\, clumps.  b) Fitted $\Delta v$ in Oph B. Colour scale is in \kms. c) $T_k$ in Oph B. Greyscale from 10\,K to 20\,K.  d) Ratio of the non-thermal to thermal line width components. Greyscale from $\sigma_{NT}\,/\,c_s = 0$ to 2.4. e) Total column density of H$_2$ derived from 850\,\micron\, dust continuum observations in Oph B, regridded to match the combined \amm\, observations. The dust temperature $T_d$ per pixel was assumed to be equal to the gas temperature $T_k$ derived from HFS line fitting of the \amm\, observations. The $N$(H$_2$) values shown have been divided by $10^{21}$. f) Fractional abundance of \amm\, divided by $10^{-9}$.}
\label{fig:b-all}
\end{figure}

\begin{figure}
\plotone{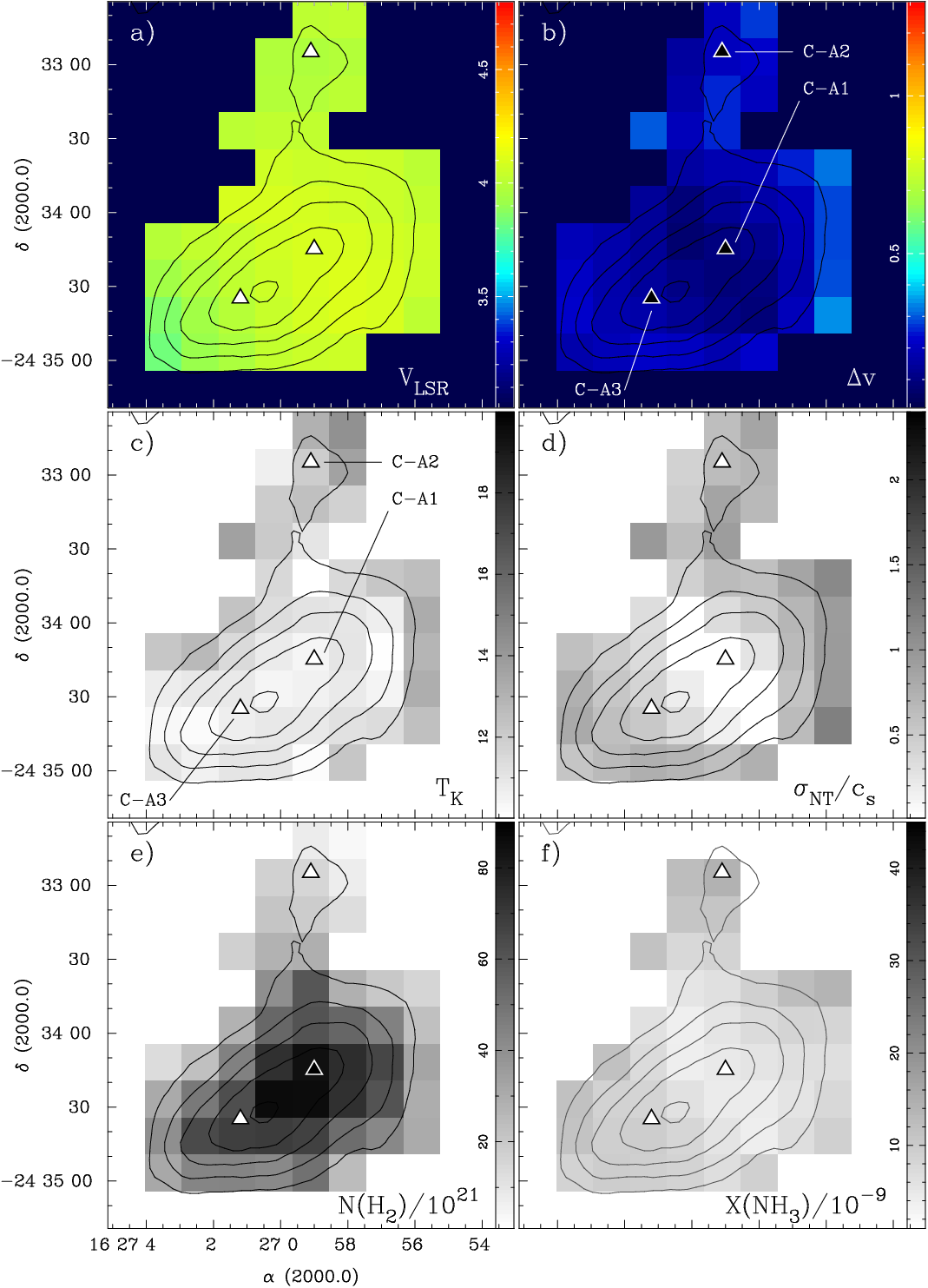}
\caption{a) Line velocity or $v_{LSR}$ in Oph C. Colour scale is in \kms. The values shown are those derived after convolving the data to 15\arcsec\, resolution and then regridding into 15\arcsec\, $\times$ 15\arcsec\, pixels. In all plots, contours show integrated \amm\, (1,1) intensity convolved to 15\arcsec\, resolution, beginning at 4.5\,K\,\kms\, and increasing by 1.5\,K\,\kms. Stars indicate protostar positions and triangles indicate positions of \amm\, clumps. b) Fitted $\Delta v$ in Oph C. Colour scale is in \kms. c) $T_k$ in Oph C. Greyscale from 10\,K to 20\,K. d) Ratio of the non-thermal to thermal line width components. Greyscale from $\sigma_{NT}\,/\,c_s = 0$ to 2.4. e) Total column density of H$_2$ derived from 850\,\micron\, dust continuum observations in Oph C, convolved to 15\arcsec\, resolution and regridded to match the combined \amm\, observations. The dust temperature $T_d$ per pixel was assumed to be equal to the gas temperature $T_K$ derived from HFS line fitting of the \amm\, observations. The $N$(H$_2$) values shown have been divided by $10^{21}$.  f) Fractional abundance of \amm\, divided by $10^{-9}$. }
\label{fig:c-all}
\end{figure}

\begin{figure}
\epsscale{0.85}
\plotone{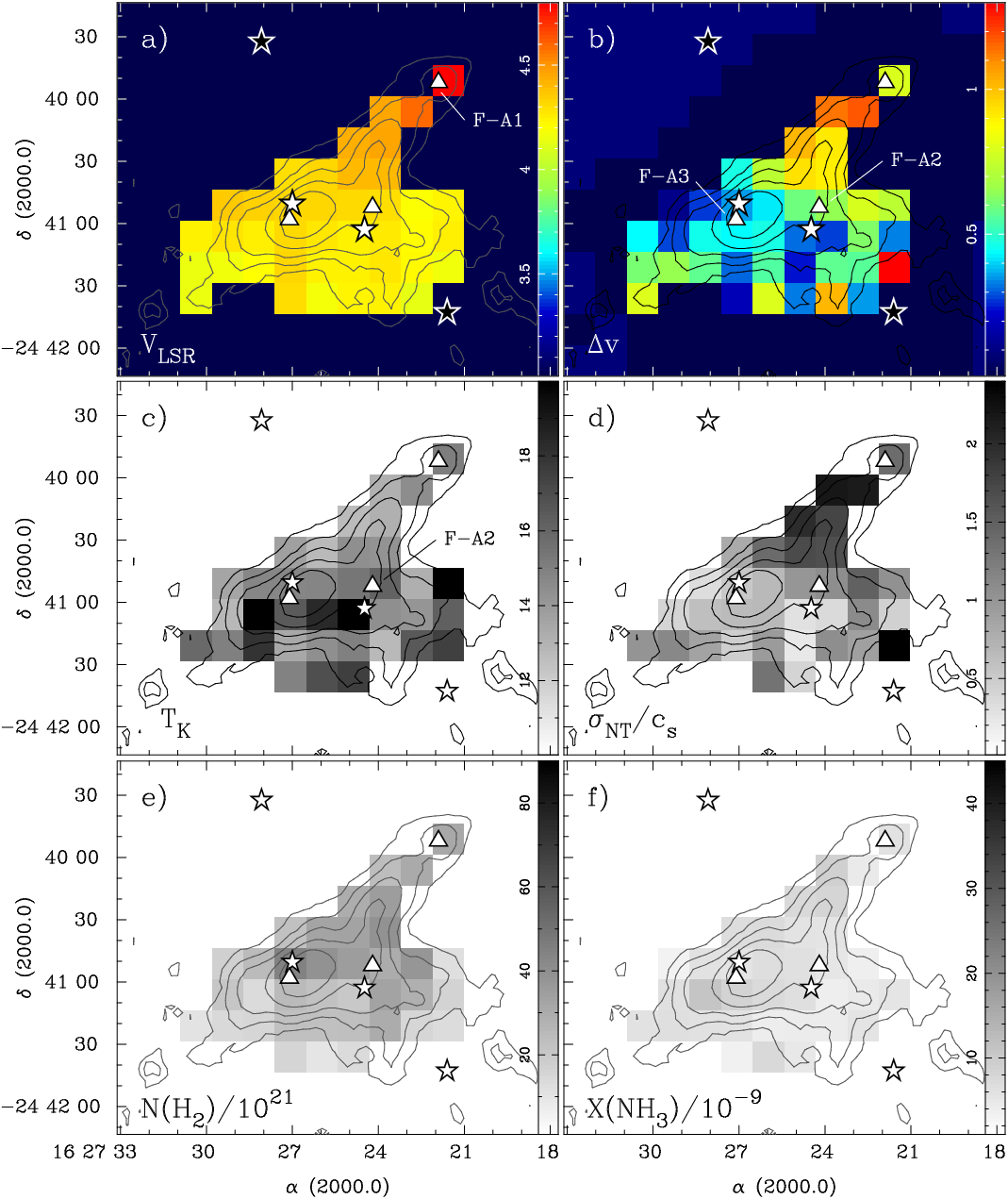}
\caption{a) Line velocity or $v_{LSR}$ in Oph F. Colour scale is in \kms. The values shown are those derived after convolving the data to 15\arcsec\, resolution and then regridding into 15\arcsec\, $\times$ 15\arcsec\, pixels. In all plots, contours show integrated \amm\, (1,1) intensity convolved to 15\arcsec\, resolution, beginning at 3\,K\,\kms\, and increasing by 1\,K\,\kms. Stars indicate protostar positions and triangles indicate positions of \amm\, clumps. b) Fitted $\Delta v$ in Oph F. Colour scale is in \kms. c) $T_k$ in Oph C. Greyscale from 10\,K to 20\,K. d) Ratio of the non-thermal to thermal line width components. Greyscale from $\sigma_{NT}\,/\,c_s = 0$ to 2.4. e) Total column density of H$_2$ derived from 850\,\micron\, dust continuum observations in Oph F, convolved to 15\arcsec\, resolution and regridded to match the combined \amm\, observations. The dust temperature $T_d$ per pixel was assumed to be equal to the gas temperature $T_k$ derived from HFS line fitting of the \amm\, observations. The $N$(H$_2$) values shown have been divided by $10^{21}$. f) Fractional abundance of \amm\, divided by $10^{-9}$. }
\label{fig:f-all}
\epsscale{1}
\end{figure}

Figures \ref{fig:b-all}a, \ref{fig:c-all}a, and \ref{fig:f-all}a show maps of $v_{LSR}$ of the fitted  \amm\, (1,1) line in Oph B, C and F respectively from the combined, smoothed and regridded data. These maps reveal that although variations of $v_{LSR}$ are seen within the Cores, they are not that kinematically distinct from each other. For example, only $0.35$\,\kms\, ($\lesssim$ the mean $\Delta v$) separates the average line-of-sight velocity in Oph B from Oph F. This result agrees with the 1D velocity dispersion of $\sim 0.36$\,\kms\, found by \citet{andre07} through \dia\, observations of the Oph cores.

In Oph B (see Figure \ref{fig:b-all}a), the $v_{LSR}$ of \amm\, emission has little internal variation, with a mean $v_{LSR} = 3.96$\,\kms\, and an rms of only $0.24$\,\kms. An overall gradient is seen across Oph B1 and B2, with smaller $v_{LSR}$ values (3.2\,\kms) at the southwest edge of B1 increasing to 4.6\,\kms\, at the most eastern part of B2. Correspondingly, B1 has a characteristic velocity somewhat less than the average ($3.8\mbox{\,\kms\,} \pm 0.1$\,\kms), while the B2 $v_{LSR}$ is slightly greater ($4.1 \pm 0.2$\,\kms). The filament connecting B1 and B2 is kinematically more similar to B2, but there is no visible discontinuity in line-of-sight velocity of the lines. B3 has the lowest $v_{LSR}$ in Oph B, with an average velocity of $3.3\mbox{\,\kms\,} \pm 0.2$\,\kms, or $0.7$\,\kms\, less than the average of the group. This large difference, greater than that between the mean $v_{LSR}$ values of Oph B and F, suggests that B3 may not be at the same physical distance as the rest of Oph B. The change in velocity between B2 and B3 occurs over a small projected distance ($\sim 30$\arcsec, or $\sim 3600$\,AU). There is some indication that B2 and B3 may overlap along the line of sight, as the determined $v_{LSR}$ values in B2 immediately south of B3 are less than those to the east and west, as might be expected if lower velocity emission is also contributing to the line at that location. (As discussed further below, the $\Delta v$ line widths in this area are larger than the average, as would be expected if unresolved emission from two different velocities is contributing to the observed line.) 

Oph C (see Figure \ref{fig:c-all}a) has a mean $v_{LSR} = 4.01$\,\kms\, with an rms of only $0.07$\,\kms. A small velocity gradient of $\sim 0.4$\,\kms\, is evident in Oph C, with a minimum line-of-sight velocity of $\sim 3.7$\,\kms\, in the southeast, increasing to a maximum of $\sim 4.1$\,\kms\, in the northwest. A similar gradient was noted in \dia (1-0) observations by \citet{andre07}, and may be indicative of rotation. The identification of two \amm\, clumps in the region, however, could also be indicative of two objects with slightly different $v_{LSR}$. A slight decrease in $v_{LSR}$ values is seen in the \amm\, extension to the north, with C-N2 associated with emission at a slightly lower $v_{LSR}$ than the mean.

Oph F (see Figure \ref{fig:f-all}a) has a mean $v_{LSR} = 4.34$\,\kms\, and an rms of only $0.26$\,\kms\, over the region containing the bulk of the \amm\, (1,1) emission and two of the three identified \amm\, peaks, despite the presence of four protostars. The filament extending towards the northwestern \amm\, peak gradually increases in $v_{LSR}$, but only by $\sim 0.2$\,\kms. With higher sensitivity, single-dish data show that outside this area $v_{LSR}$ drops to $\sim 3.7$\,\kms. There is clear evidence for two velocity components along the line of sight at the peak position of F-A1, with a secondary component at $\sim 3.7$\,\kms. \citet{andre07} also find two velocity components near F-A1 in \dia observations. For the brighter component, their \dia\, data agree with our \amm\, data, but for the secondary component they find a higher $v_{LSR} = 4.1$\,\kms. Some blue asymmetry is found in the line profiles of F-A2 and F-A3, but this is more likely due to the complicated velocity structure of the core rather than infall motions (comparison with an optically thin tracer at this position is necessary to confirm infall).

In summary, $v_{LSR}$ varies little across any of the Cores (rms $ < 0.24$\,\kms), and additionally varies little between them (maximum mean difference $\sim 0.38$\,\kms). Some small gradients were found in the larger Cores (i.e., on scales larger than the individual \amm\, clumps) which may be indicative of rotation. 

Figures \ref{fig:b-all}b, \ref{fig:c-all}b, and \ref{fig:f-all}b show the line $\Delta v$ for Oph B, C and F obtained from the combined, smoothed and regridded data. As stated above, the line widths have been corrected for the resolution of the spectrometer (0.3\,\kms). In the few cases where the returned FWHM from the fits is similar or equal to the resolution, we set the corrected FWHM to the thermal line width (see Appendix A for further discussion). 

Line widths range from $\Delta v \lesssim 0.1$\,\kms\, to $\sim1.2$\,\kms\, in both Oph B and F, with a minimum line width of 0.08\,\kms\, in B3 and a maximum of 1.37\,\kms\, in B2. Line widths in Oph C range from 0.11\,\kms\, to 0.70\,\kms. The rms variations $\Delta v$ and $v_{LSR}$ in Oph B and C are similar (0.2\,\kms\, and 0.1\,\kms, respectively), while in Oph F the rms in $\Delta v$ is much larger than the rms in $v_{LSR}$ (i.e., 0.3\,\kms\, compared to 0.12\,\kms, respectively). 

The extended emission in Oph B is dominated by highly non-thermal motions (mean $\Delta v = 0.83$\,\kms), shown in Figure \ref{fig:b-all}b. Several localized pockets of narrow line width are found embedded within the more turbulent gas. The single Oph B3 clump (B3-A1) and B2-A7 are both characterized by extremely narrow $\Delta v$ (i.e., 0.08\,\kms\, and 0.33\,\kms\, respectively).  As mentioned above, the line emission broadens from B3 to B2 over only $\sim 30$\arcsec\, (3600\,AU) to $\Delta v \, \sim 1.4$\,\kms, possibly due to line blending along the line of sight if B2 and B3 overlap in projection at these locations. We find small $\Delta v \sim 0.5$\,\kms\, towards the south-eastern edge of the mapped region in B1 near \amm\, clumps B1-A3 and B1-A4, but the line width minimum does not coincide with either core. Another region of low $\Delta v \sim 0.6$\,\kms\, is coincident with the \amm\, clumps B2-A1 and B2-A2. A final $\Delta v$ minimum, $\Delta v \sim 0.4$\,\kms\, is found between the two eastern protostars in B2. Note that the protostars in Oph B are all associated with smaller $\Delta v$ than the average value for the core, but none are coincident with a clear local minimum in line width. 

The maximum line width in Oph C, $\Delta v = 0.70$\,\kms,  is half that found in Oph B. Figure \ref{fig:c-all}b shows that most of the emission in C is narrow, with a minimum $\Delta v  = 0.11$\,\kms, similar to the extremely narrow lines found in B3. These narrowest line widths are found centered on \amm\, clump C-A2 in a band perpendicular to the elongated direction of Oph C and are coincident with the highest velocity emission, and are thus not coincident with the \amm\, integrated intensity maximum nor the continuum emission peak. Curiously, if C-A2 and C-A3 are indeed physically distinct clumps, we would expect the broadest lines between them due to overlap, but instead find narrow lines at this location. The largest line widths are found at the edges of the integrated intensity contours. 

Oph F is characterized by moderately wide line emission more similar to that found in Oph B, with a mean $\Delta v = 0.63$\,\kms\, (see Figure \ref{fig:f-all}b). Line widths associated with the central \amm\, clumps F-A2 and F-A3 are smaller than the mean (i.e., 0.6\,\kms\, and 0.3\,\kms\, respectively), while clear minima in line width ($\Delta v \sim 0.33$\,\kms) are found at the locations of the two central protostars. Line widths along the Core extending to the northwest integrated intensity peak broaden to larger values ($\sim 1.2$\,\kms), but at the tip \amm\, clump F-A1 is associated with $\Delta v = 0.4$\,\kms\, in a single 15\arcsec\, pixel. 

Overall, the observed \amm\, $\Delta v$ in the Cores are generally large, excepting Oph C. We also find regions of localized narrow line emission, some of which are associated with \amm\, clumps. Line widths near protostars tend to be smaller than the mean values, but only in Oph F are the protostars coincident with clear minima in $\Delta v$.

\subsection{Kinetic Temperatures and Non-Thermal Line widths}

Following \citet{mangum92}, among others, we use the returned $\tau$, $\Delta v$ and line brightnesses of the \amm\, (1,1) and (2,2) lines in each pixel to calculate the kinetic temperature of the gas.  Details of our calculations can be found in Appendix \ref{ap_line}. Propagating uncertainties from our hyperfine structure fitting routine gives typical uncertainties in $T_K$ of $\sim 1$\,K. The mean, rms, minimum and maximum kinetic temperatures are given for Oph B, C and F in Table \ref{tab:tketc}. Figures \ref{fig:b-all}c, \ref{fig:c-all}c and \ref{fig:f-all}c show the kinetic temperatures calculated across the Oph Cores. 

In Oph B, we find a mean $T_K = 15.1$\,K with an rms variation across the entire core of only 1.8\,K. Most \amm\, emission peaks are associated with lower than average gas temperatures, but only a few are coincident with clear $T_K$ minima. The lowest temperatures in the Core, $T_K = 13.7$\,K and 12.4\,K, are found in southern B1 towards B1-A3 and B1-A4, respectively. These low temperatures are found at the same location as the detection of single dish C$_2$S emission. Gas temperatures appear colder ($T_K \sim 13$\,K) towards the centre of Oph B2, but the minimum temperature, $T_K = 12.4$\,K, is not coincident with an \amm\, clump. Instead, the lowest temperatures are found directly between the \amm\, clumps B2-A4, B2-A5 and B2-A6, and closer to the central submillimeter clump. 

Oph C is the coldest of the observed cores, with a mean $T_K = 12.8$\,K and a similar rms variation (1.6\,K) as in Oph B. The central region is effectively at a single low temperature $T_K = 10.6$\,K, with the lowest values found near the emission peaks C-A1 and C-A3 in the northwest and southeast, while the gas temperature of the northern core, C-A2, is consistent with the average. 

Oph F is characterized by the highest temperatures of the observed Cores, with a mean $T_K = 16.6$\,K, slightly warmer than Oph B, and with a large rms variation of 3.2\,K. No clear minima in gas temperature are observed near any of the \amm\, clumps, protostars or continuum peaks identified in the region. The protostar associated with \amm\, clump F-A2 is coincident with a temperature maximum in a single 15\arcsec\, pixel. 

The gas temperatures traced by \amm\, emission in the Oph B and F Cores are consistently higher than those found in isolated dense clumps. For example, all five of the starless clumps surveyed by \citet{tafalla02} were found to have a constant gas temperature $T_K = 10$\,K determined through an analysis similar to that done here.  Two recent studies of \amm\, emission in dense clumps in the Perseus molecular cloud and the less active Pipe Nebula at 32\arcsec\, resolution also found slightly lower temperatures than those found here, with a median $T_K = 11$\,K in Perseus \citep{rosolowsky08a} and a mean $T_K = 13$\,K $\pm 3$\,K for $\lesssim 1$\,M$_\odot$ clumps in the Pipe Nebula \citep[][we note that one object in the Pipe is warmer than the typical $T_K$ found in Oph B and F]{rathborne08}. In a sample of \amm\, observations towards Galactic high mass star forming regions, \citet{wu06} found a mean $T_K = 19$\,K. The mean kinetic temperatures found in Oph B and F are also slightly greater than the median $T_K = 14.7$\,K found in a survey of \amm\, observations by \citet{jijina99}, although their analysis showed that the median temperature of dense gas in clusters was significantly higher, $T_K = 20.5$\,K, than in non-clustered environments where the median $T_K = 12.4$\,K. Since L1688 is a clustered star forming environment, it is not unreasonable to expect temperatures higher than those in more isolated regions, given the \citeauthor{jijina99} results.  

Evidence for an extremely cold temperature of 6\,K [obtained through observations of H$_2$D$^+\, (1_{10} - 1_{11})$, which likely probes denser gas than \amm\, (1,1) and (2,2), was recently found for the nearby Oph D Core \citep{harju08}. In addition, while the mean gas temperature $T_K = 12$\,K in Oph C, temperatures in the highest column density gas drop to 10\,K, similar to temperatures found in the studies of isolated clumps described above. 

We find little difference in average gas temperatures between \amm\, clumps and \amm\, emission associated with submillimeter continuum emission peaks, and a mean $T_K$ increase of only $\sim 1$\,K, i.e., similar to our uncertainty in $T_K$, in gas temperatures near protostars (but note only five protostars are associated with emission with sufficient S/N to fit the HFS). The \citet{jijina99} survey found that \amm\, cores not associated with identified IRAS sources (presumably protostellar objects) were slightly colder than those with coincident IRAS detections (12.4\,K compared with 15.0\,K), but the effect was much smaller than temperature differences seen due to association with a cluster.

Given the determined gas temperature $T_K$, we calculate the expected one-dimensional thermal velocity dispersion $\sigma_T$ of the gas across the cores: 

\begin{equation}
\sigma_T = \sqrt{\frac{k_B T_K}{\mu_{\mbox{\tiny{\amm}}} m_{\mbox{\tiny{H}}}}}
\end{equation}

Here, $k_B$ is the Boltzmann constant, $\mu_{\mbox{\tiny{\amm}}} = 17.03$ is the molecular weight of \amm\, in atomic units, and $m_{\mbox{\tiny{H}}}$ is the mass of the hydrogen atom. Similarly, the thermal sound speed $c_s$ of the gas can be calculated using a mean molecular weight $\mu = 2.33$. 

The non-thermal velocity dispersion $\sigma_{NT}$ is given by  

\begin{equation}
\sigma_{NT} = \sqrt{\sigma^2_{obs} - \sigma^2_T}
\end{equation}

\noindent where $\sigma_{obs} = \Delta v / (2 \sqrt{2 \ln{2}})$. The mean, rms, minimum and maximum values for both $\sigma_{NT}$ and the non-thermal to thermal velocity dispersion ratio of the gas, given by $\sigma_{NT}\,/\,c_s$, are given for each of Oph B, C and F in Table \ref{tab:tketc}. 

Figures \ref{fig:b-all}d, \ref{fig:c-all}d and \ref{fig:f-all}d show the resulting non-thermal to thermal velocity dispersion ratio over the cores. The mean $\sigma_{NT}\,/\,c_s$ values show supersonic velocities are present. At the limits of the velocity resolution of our data, the smallest observed line widths are consistent with motions being purely thermal in nature. 

We find that the mean $\sigma_{NT} = 0.35$\,\kms\, and $\sigma_{NT}\,/\,c_s = 1.5$ in Oph B. Across the Core, we find a moderate $\sigma_{NT}\,/\,c_s$ rms of 0.4. The majority of the gas traced by \amm\, in Oph B is thus dominated by non-thermal, mildly supersonic motions.  Several \amm\, clumps are associated with smaller, but still transsonic, non-thermal motions. Thermal motions dominate the observed line widths in only two well-defined locations in Oph B which additionally coincide with \amm\, clumps: B2-A7, with $\sigma_{NT} / c_s = 0.5$, and B3-A1, where the observed line width is consistent with purely thermal motions. Otherwise, little difference is seen between the non-thermal line widths of individual cores and the surrounding gas, with a mean $\sigma_{NT}\,/\,c_s = 1.4$\,\kms\, for the \amm\, clumps. 


In contrast, Oph C has a mean $\sigma_{NT} = 0.14$\,\kms. Consequently, the mean $\sigma_{NT}\,/\,c_s = 0.6$ with an rms of only 0.2, showing that thermal motions dominate the observed line widths over much of the Core. The minimum non-thermal line width is found associated with C-A1, which is consistent with purely thermal motions within our velocity resolution limits. On average, non-thermal motions in Oph F are also similar to the expected thermal values, with a mean $\sigma_{NT}\,/\,c_s = 0.9$, but with a larger spread around the mean (0.6 rms) and a maximum value ($\sigma_{NT}\,/\,c_s = 2.2$) similar to that found in Oph B ($\sigma_{NT}\,/\,c_s = 2.5$). The lowest values are found towards F-A2 and F-A3 and the nearby protostars. 

The non-thermal \amm\, line widths we measure are similar to those recently found for \dia (1-0) emission in the Cores at larger angular resolution ($\sim 26$\arcsec), where the mean $\sigma_{NT}/ c_s = 1.6 \pm 0.3$ in Oph B, $0.9 \pm 0.2$ in Oph C, $1.5 \pm 0.8$ in Oph F \citep{andre07}, and are less than those found in DCO$^+$ emission ($\sigma_{NT} / c_s \sim 2$ in B2, $\sim 1 - 1.5$ in B1, B3, C and F \citep{loren90}. 

\subsection{Column Density and Fractional Abundance}

\subsubsection{\amm}

Given $\Delta v$, $\tau$ and $T_{ex}$ from the \amm\, (1,1) line fitting results, we calculate the column density of the upper level of the \amm\, (1,1) inversion transition. We then calculate the \amm\, partition function (given $T_K$) to determine the total column density of \amm\, following \citet{rosolowsky08a}. Relevant equations are given in Appendix \ref{ap_line}. 

We also calculate the H$_2$ column density, $N(\mbox{H}_2)$, per pixel in the Cores from 850\,\micron\, continuum data using

\begin{equation}
\label{eqn:dust}
N(\mbox{H}_2) = S_\nu / [ \Omega_m \mu m_{\mbox{{\tiny H}}} \kappa_\nu B_\nu (T_d)] ,
\end{equation}

\noindent where $S_\nu$ is the 850\,\micron\, flux density, $\Omega_m$ is the main-beam solid angle, $\mu = 2.33$ is the mean molecular weight, $m_{\mbox{{\tiny H}}}$ is the mass of hydrogen, $\kappa_\nu$ is the dust opacity per unit mass at 850\,\micron, and $B_\nu (T_d)$ is the Planck function at the dust temperature, $T_d$. We take $\kappa_\nu = 0.018$\,cm$^2$\,g$^{-1}$, following \citet{shirley00}, using the dust model from \citet{ossen94} which describes grains that have coagulated for $10^5$\,years at a density of $10^6$\,\cc\, with accreted ice mantles and incorporating a gas-to-dust mass ratio of 100. The 15\arcsec\, resolution continuum data were regridded to 15\arcsec\, pixels to match the combined \amm\, observations. The dust temperature $T_d$ per pixel was assumed to be equal to the gas temperature $T_k$ derived from HFS line fitting of the combined \amm\, observations. This assumption is expected to be good at the densities probed by our \amm\, data ($n \gtrsim 10^4$\cc), when thermal coupling between the gas and dust by collisions is expected to begin \citep{goldsmith78}. If these temperatures are systematically high, however, then the derived $N(\mbox{H}_2)$ values are systematically low. There is a $\sim 20$\% uncertainty in the continuum flux values, and estimates of $\kappa_\nu$ can additionally vary by $\sim 3$ \citep{shirley00}. Our derived $H_2$ column densities consequently have uncertainties of factors of a few. 

Due to the chopping technique used in the submillimeter observations to remove the bright submillimeter sky, any large scale cloud emission is necessarily removed. As a result, the image reconstruction technique produces negative features around strong emission sources, such as Oph B \citep{johnstone00b}. While the flux density measurements of bright sources are likely accurate, emission at the core edges underestimates the true column. We thus limit our analysis to pixels where $S_\nu \geq 0.1$\,Jy\,beam$^{-1}$, though the rms noise level of the continuum map is $\sim 0.03$\,Jy\,beam$^{-1}$. For a dust temperature $T_d = 15$\,K, this flux level corresponds to $N(\mbox{H}_2) \sim 6 \times 10^{21}$\,cm$^{-2}$. 


Using the calculated H$_2$ and \amm\, column densities, we have calculated per pixel the fractional abundance of \amm\, relative to H$_2$, $X$(\amm) $=  N(\mbox{\amm})\,/\,N(\mbox{H}_2)$ for each Core. The results of these calculations are shown in Figures \ref{fig:b-all}, \ref{fig:c-all} and \ref{fig:f-all}, which show the H$_2$ column density derived from submillimeter continuum data, $N(\mbox{H$_2$})$ and the fractional \amm\, abundance, $X(\mbox{NH$_3$})$. The mean, rms, minimum and maximum of the derived column density and fractional abundance in each Core are given in Table \ref{tab:tketc}, while specific values for identified \amm\, clumps are listed in Table \ref{tab:peak_dat}. The $N(\mbox{H$_2$})$ and consequently the $X(\mbox{\amm})$ uncertainties given in Table \ref{tab:peak_dat} include the $\sim 20$\% uncertainty in the submillimeter continuum flux values only; uncertainty in $\kappa_\nu$ is not taken into account. 

Oph B has a mean \amm\, column density of $2.2 \times 10^{14}$\,cm$^{-2}$, with the highest $N(\mbox{\amm})$ values (maximum $N(\mbox{\amm}) = 4.8 \times 10^{14}$\,cm$^{-2}$) found in B1. Two peaks in \amm\, column density are found in B1 which correspond closely with the integrated intensity maxima, but are offset from the B1 \amm\, clumps. In B2, the column density also generally follows the integrated intensity contours, with lower values overall than in B1. The highest \amm\, column in B2 of $N(\mbox{\amm}) = 4 \times 10^{14}$\,cm$^{-2}$ is found towards B2-A5. The highest opacity in B2, $\tau = 4.7$, was found associated with B2-A7, but the \amm\, column density at this location is similar to the core average. Small column density increases are seen at other \amm\, intensity peak locations. The \amm\, extension connecting B1 and B2 is characterized by similar \amm\, column densities to those at the edges of the Core with no obvious $N(\mbox{\amm})$ maxima. 

The discrepancy between the bright \amm\, and faint submillimeter continuum emission in B1 indicates a high \amm\, fractional abundance relative to B2, with fractional abundances a factor of $\gtrsim 2-3$ higher than the typical values of $X(\mbox{\amm}) \sim 10^{-8}$ within B2. This is shown in Figure \ref{fig:b-all}f.  Abundance {\it minima} are seen in B2, most notably towards B2-A6 and nearby protostars. Despite the higher \amm\, column densities in B1 and B2, the lack of submillimeter emission in the \amm\, extension connecting the two regions suggests this connecting material has a higher fractional \amm\, abundance. Prominent negative features in the continuum data in this extension preclude a quantitative $X(\mbox{\amm})$ estimate. 

The mean and maximum \amm\, column densities in Oph C are similar to those found in Oph B ($\langle N(\mbox{\amm}) \rangle = 2.4 \times 10^{14}$\,cm$^{-2}$ and $N(\mbox{\amm}) = 5.5 \times 10^{14}$\,cm$^{-2}$, respectively). Oph C contains two $N(\mbox{\amm})$ maxima. One is coincident with C-A3 and the second is offset to the west by $\sim 15$\arcsec\, from C-A1. C-A3 is correspondingly associated with a maximum in \amm\, fractional abundance ($X(\mbox{\amm}) = 8.8 \times 10^{-9}$), but C-A1 is coincident with an elongated minimum $X(\mbox{\amm}) \sim 3 \times 10^{-9}$ that extends along the same axis perpendicular to the long axis of the core where the smallest line widths were found. The highest \amm\, abundances $X(\mbox{\amm}) \sim 12 \times 10^{-9}$, are found in the northern extension. The mean abundance in Oph C, $X(\mbox{\amm}) = 8.2 \times 10^{-9}$, is slightly more than half the Oph B average. 


In Oph F, the mean $N(\mbox{\amm}) = 1.4 \times 10^{14}$\,cm$^{-2}$ is less than that found in Oph B and C by a factor of $\sim 2$. The maximum $N(\mbox{\amm}) = 2.5 \times 10^{14}$\,cm$^{-2}$ is also significantly less than the maxima in either B or C, and is found in the emission extending to the northwest from the two central \amm\, clumps and protostars. The fractional \amm\, abundances are also low compared with B and C, with a mean $X(\mbox{\amm}) = 5.4 \times 10^{-9}$ and a maximum $X(\mbox{\amm}) = 1.0 \times 10^{-8}$ found near but not coincident with F-A3.  

Studies of isolated starless clumps have determined a wide range of \amm\, abundance values for these objects. While in some cases different methods have been used to determine H$_2$ column density values than that performed here, `typical' observed fractional abundance values in cold, dense regions tend to be on the order of a few $\times 10^{-9}$ to a few $\times 10^{-8}$ \citep{tafalla06,crapsi07,ohishi92,larsson03,hotzel01}. Abundances as low as $X(\mbox{\amm}) = 7 \times 10^{-10}$  and $8.5 \times 10^{-10}$ have been proposed for B68 \citep{difran02} and Oph A \citep{liseau03}, respectively. The values found here agree well with previous studies. The wide variations of $X(\mbox{\amm})$ in the same general environment suggests dramatic differences in the chemical states of the Cores in L1688 (see \S 5). 

\subsubsection{C$_2$S and HC$_5$N}

Similarly, we can calculate the abundance of C$_2$S and HC$_5$N from respective emission detected in the single-dish data, where

\begin{equation}
N = \frac{8 \pi k \nu_0}{h c^2} \frac{g1}{g2} \frac{1}{A_{ul}} \sqrt{2 \pi} \sigma_v [J(T_{ex}) - J(T_{bg})] \tau_{ul}
\end{equation}

\noindent is the column density of the upper state of the observed transition \citep{rosolowsky08a}. The values for $A_{ul}$, $g_1$ and $g_2$ were taken from \citet{pickett98} for each transition. Assuming the transitions are optically thin, the observed temperature of the line $T_{MB} = [J(T_{ex}) - J(T_{bg})] \tau_{ul}$. The partition function $Z = \sum_i g_i \exp(\frac{-E_i}{kT_K})$ was then used to calculate the total column density of each species as for \amm, with $g_i$ and $E_i$ values taken from \citet{pickett98}. The column densities thus derived are given in Table \ref{tab:ccsdat}. The molecular column densities derived for C$_2$S in B1 and C ($N(\mbox{C}_2\mbox{S}) \sim 10^{12-13}$\,cm$^{-2}$) are similar to results in young starless cores \citep{tafalla06,rosolowsky08a,lai03}. The $N(\mbox{HC$_5$N})$ results agree with previous measurements in the Taurus molecular cloud \citep{codella97,bm83} and the Pipe Nebula \citep{rathborne08}. We calculate molecular abundances as above and find $X(\mbox{C$_2$S}) = 3.1 \times 10^{-10}$ and $1.5 \times 10^{-10}$ at the C$_2$S emission peaks in Oph B1 and C, respectively. We further find an abundance $X(\mbox{HC$_5$N}) = 4.6 \times 10^{-11}$ at the HC$_5$N emission peak in Oph C. 

\subsection{H$_2$ Density}

Given the determined excitation and kinetic temperatures, $T_{ex}$ and $T_K$, and assuming the metastable states can be approximated as a two level system, we have calculated the gas density $n(\mbox{H$_2$})$ from the \amm\, (1,1) transition following \citet{hotownes}. We list the mean, rms variation and range of densities found for each core in Table \ref{tab:tketc}. Note that this density is effectively a mean density along the line of sight. In general, we find $n(\mbox{H$_2$}) \sim$ a few $\times 10^4$\,\cc\, in all three Cores, with only moderate variation and no clear spatial correspondence with \amm\, or continuum clumps. The largest $n(\mbox{H$_2$})$ values ($n(\mbox{H$_2$}) \sim 8 \times 10^5$\,\cc) were found in Oph B2 towards the central continuum clump MM8 (labelled in Figure \ref{fig:b-compare}b). While these values agree with $n(\mbox{H$_2$})$ estimates based on \amm\, emission in other regions, they are an order of magnitude lower than estimates of Ophiuchus clump densities derived from dust continuum emission studies at similar spatial resolutions \citep{motte98,johnstone00}.

\section{Discussion}

\subsection{Discussion of small-scale features}

\subsubsection{Correlation between \amm\, clumps, \amm\, integrated intensity and dust clumps}

In \S 3, we used {\sc clumpfind} to identify objects in \amm\, emission within the Oph Cores in position and velocity space. The returned \amm\, clump locations are generally found at locations of peak integrated \amm\, intensity, with the exception of Oph C, in which we found two distinct \amm\, clumps offset from the \amm\, integrated intensity maximum. 

In Oph B, we find poor correlation between maxima of \amm\, integrated intensity and thermal dust continuum emission. Since most \amm\, clumps are located at integrated intensity maxima, we hence find \amm\, clumps identified through {\sc clumpfind} do not correlate well with dust clumps. Continuum dust emission is a commonly used surrogate tracer of gas column density. The observed flux is a function of the dust emissivity ($\kappa_\nu$) and temperature ($T_d$). If the dust in B1, for example, was colder than that in B2, the same column of dust would produce less emission. In \S 4.3, we calculated H$_2$ column densities assuming the dust and gas are thermally coupled. For Oph B, we found the H$_2$ column density closely followed the observed continuum emission under this assumption (see Figure \ref{fig:b-all}a vs. Figure \ref{fig:b-compare}b). The dust and gas, however, may not have the same temperature. If the dust is colder than the gas by a small amount ($T_d = 10$\,K compared with $T_K = 15$\,K, for example), the true column density of H$_2$ could be larger by a factor of $\sim 2$ along that line of sight. If these temperature differences occur on small enough scales, e.g., at the \amm\, clump positions, they could explain the discrepancy between the locations of dust clumps and \amm\, clumps. Thermal coupling of gas and dust is most likely to occur, however, in the coldest and densest clumps, i.e., exactly where we do not find correspondence between the dust and gas tracers. 


A more likely cause of the offset between dust and \amm\, emission is fractional abundance variation of \amm\, in the Oph Cores. If the column densities determined from the dust emission are accurate, then most dust clumps are associated with $X(\mbox{\amm})$ minima. Within B1 and B2, we find variations in $X(\mbox{\amm})$ of $\gtrsim 2$ on length scales similar to the \amm\, clump sizes.  Models of nitrogen chemistry in dense regions suggest that a long timescale, greater than the free-fall time, is required for molecules such as \amm\, to achieve steady state values, but that during gravitational collapse $X(\mbox{\amm})$ begins to decrease at densities $n \gtrsim 10^6$\,\cc\, \citep{aikawa05,flower06}. 

The C$_2$S molecule is easily depleted in cold, dense environments with an estimated lifetime of a few $\times 10^4$\,yr \citep{gregmon06}. It is thus a good tracer of young, undepleted cores \citep{suzuki92,lai00,tafalla04}. The detection of C$_2$S in southern Oph B1 and Oph C therefore suggests that these specific locations are chemically, and hence dynamically, younger compared with other regions. The C$_2$S emission detected in both B1 and C is coincident with or only slightly offset from the integrated \amm\, intensity peak, suggesting significant depletion has not yet occurred at those particular locations. This conclusion is further bolstered by the fact that we find higher gas densities (see \S 4.5) in B2 than in B1 or C, and both B2 and F are associated with embedded protostars and are therefore likely more dynamically evolved, i.e., denser. The higher levels of non-thermal motions found in Oph B are at odds with what is expected for an evolved, star forming core, however, and we discuss this further below. 

 \subsubsection{Comparison of \amm\, clumps, submillimeter clumps and protostars}
 
We next compare the mean properties of the dense gas associated with the locations of \amm\, clumps, submillimeter clumps and protostars. We note that given the poor correlation between the \amm\, clumps and submillimeter clumps and protostars, the derived physical properties (e.g., $T_K$ and $\sigma_{NT}$) at the submillimeter clump and protostellar locations may be associated with larger-scale gas emission along the line-of-sight rather than the dense clump gas. 

In general, we find only small differences between the mean properties of the dense gas at the peak locations of the \amm\, clumps, submillimeter clumps and embedded protostars. The mean kinetic temperatures for \amm\, clumps and submillimeter clumps are nearly equal ($\sim 14$\,K), and only $\sim 1.5$\,K less than the values associated with embedded protostars. This difference is not significant given that the uncertainties in $T_K$ are on the order of 1\,K. Conversely, excitation temperatures associated with embedded protostars are $\sim 1.5$\,K lower than that of \amm\, clumps and submillimeter clumps where $T_{ex} \sim 10$\,K, with uncertainties in $T_{ex}$ also $\sim 1$\,K.  The line widths of submillimeter clumps tend to be larger than those associated with \amm\, clumps by only $\sim 25$\%. 

Some differences in mean properties between objects are notable. For example, protostars have associated $\Delta v$ and $\sigma_{NT}\,/\,c_s$ a factor of 2 {\it narrower} than both submillimeter and \amm\, clumps. Also, submillimeter clumps and protostars have lower fractional abundances than seen for \amm\, clumps by a factor of $\sim 2$. Note, however, that only five protostars are found with \amm\, emission strong enough to analyze, as we described in \S 3. For this reason, our comparison sample is limited to protostars that are still associated with significant amounts of gas, where conditions are likely more similar to those found in submillimeter clumps than for more evolved protostars.  


\subsubsection{$\sigma_{NT}\,/\,c_s$ in Individual \amm\, Clumps}

\begin{figure}
\plotone{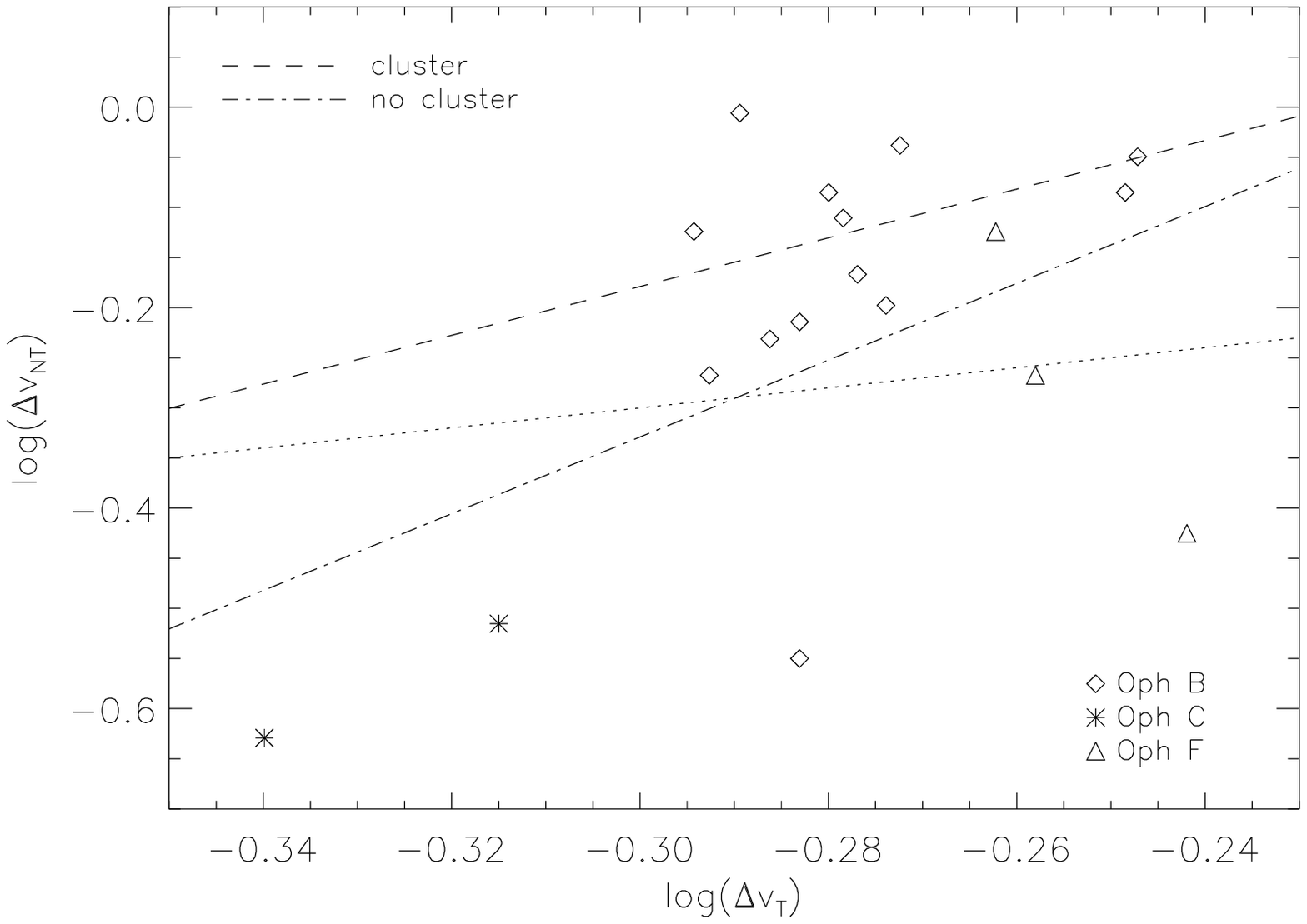} 
\caption{Non-thermal versus thermal line widths (FWHM) for individual \amm\, clumps in the Oph B, C and F Cores. The two dashed lines show relationships found by \citet{jijina99} for \amm\, clumps both with and without an associated cluster. Clumps in Oph B follow the clustered trend, but clumps in Oph C are nearer the relationship for more isolated clumps. Clumps in Oph F scatter across the plot. The dotted line represents $\Delta v_{NT} = \Delta v_T$. }
\label{fig:jijina}
\end{figure}

 We next look at the non-thermal line widths in the individual \amm\, clumps in all three cores. In general, \citet{jijina99} found that non-thermal \amm\, line widths in clustered environments are larger than those found in isolated regions. In Figure \ref{fig:jijina}, we plot $\Delta v_{NT}$ versus $\Delta v_T$ for the \amm\, clumps in each core. We omit \amm\, clumps B3-A1 and C-A1 where the corrected non-thermal line width is effectively zero. We also show the best fit lines found by \citeauthor{jijina99} to the relationship between thermal and non-thermal line widths in clustered and in isolated regions. Most of the Oph B clumps lie above the $\Delta v_{NT} - \Delta v_T$ trend for isolated clumps and near the trend for the clumps in clustered regions. The good agreement is somewhat surprising given that the majority of the objects in the \citeauthor{jijina99} sample were observed with $\sim 4 - 8$ times poorer angular resolution, while those observed with high angular resolution are high mass star forming regions $\sim 3 - 7.5$\,kpc distant and thus have very low linear resolution (excepting Orion B, at a distance of 420\,pc). Even at the small spatial scales probed by our observations, Oph B is characterized by wide line widths that follow the relationship found for larger objects in clustered environments. In comparison, the \amm\, clumps in Oph C lie well below the clustered $\Delta v_{NT} - \Delta v_T$ trend, with $\Delta v_{NT} < \Delta v_T$, and also below the $\Delta v_{NT} -\Delta v_T$ trend seen for objects not associated with a cluster. Two of the three Oph F clumps also fall significantly below the isolated object trend, while the third is more turbulent. 

\subsection{Discussion of the Cores}


\subsubsection{Trends with $N(\mbox{H$_2$})$}

\begin{figure}
\plotone{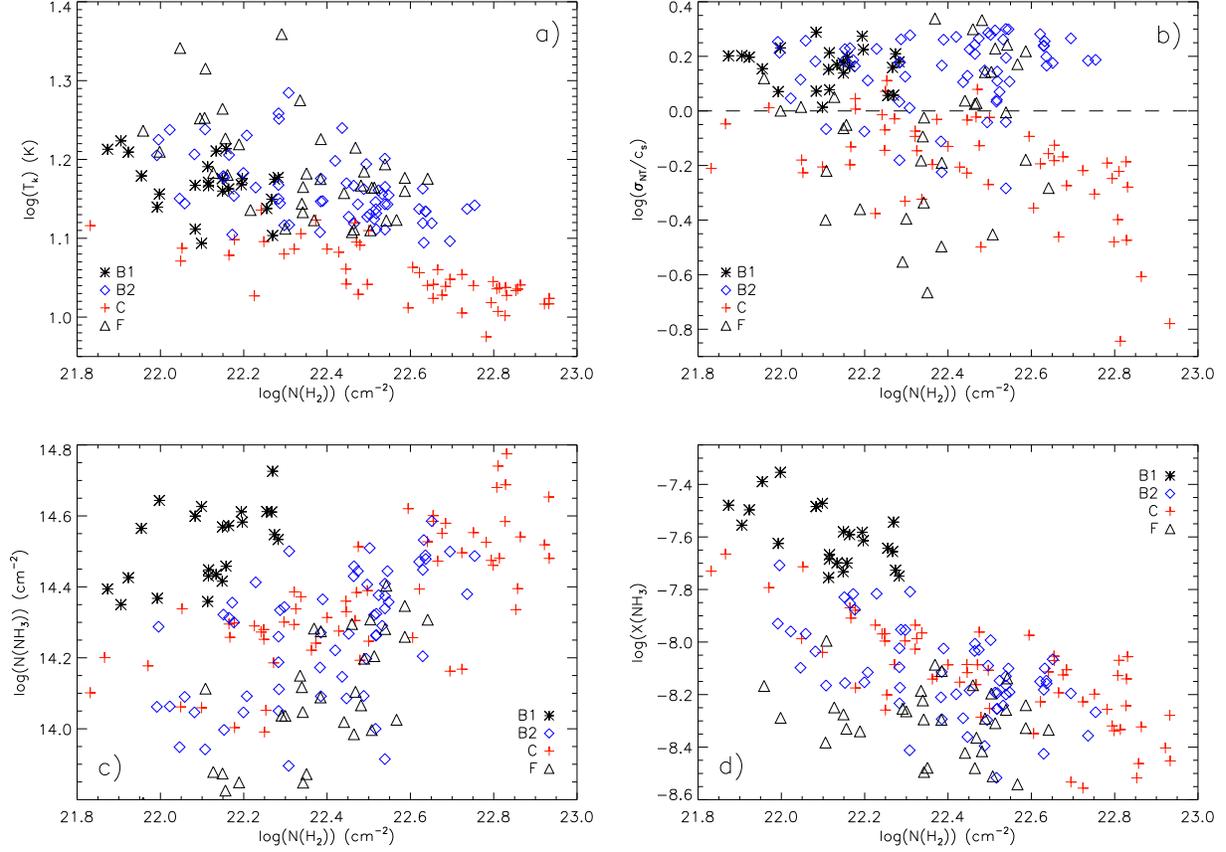} 
\caption{a) Kinetic temperature ($T_K$) in K versus $N(\mbox{H$_2$})$ (derived from the 850\,\micron\, submillimeter continuum data) in cm$^{-2}$ units for Oph B1, B2, C and F. In all plots, points shown are values in 15\arcsec\, pixels. Temperatures in Oph C are nearly universally lower than those in B1, B2 and F, and show a tendency to decrease with increasing H$_2$ column density. A significant trend in temperature with $N(\mbox{H$_2$})$ is not obvious in the other Cores. b) $\sigma_{NT}\,/\,c_s$ versus $N(\mbox{H$_2$})$ in Oph B1, B2, C and F. The dashed line indicates $\sigma_{NT}\,/\,c_s = 1$. Oph B1 and B2 are consistent with a constant ratio of non-thermal to thermal line widths, where $\sigma_{NT}\,/\,c_s > 1$, over all H$_2$ column densities. Oph C line widths are generally dominated by thermal motions, and $\sigma_{NT}\,/\,c_s$ decreases significantly at H$_2$ column densities above log($N(\mbox{H$_2$})) \sim 22.6$. c) $N(\mbox{\amm})$ (derived from the \amm\, HFS line fitting results) versus $N(\mbox{H$_2$})$ for Oph B1, B2, C and F. \amm\, column densities tend to increase with H$_2$ column densities in all Cores. Oph B1 has significantly higher $N(\mbox{\amm})$ for its relatively low $N(\mbox{H$_2$})$ values compared with the other Cores. d) $X(\mbox{\amm})$ versus $N(\mbox{H$_2$})$ for Oph B1, B2, C and F. While c) shows the \amm\, column densities tend to increase with increasing $N(\mbox{H$_2$})$, the fractional \amm\, abundances appear to decrease with higher H$_2$ column densities.  }
\label{fig:trends}
\end{figure}

In Figure \ref{fig:trends}, we plot the distribution of $T_K$, $\sigma_{NT}\,/\,c_s$, $N(\mbox{\amm})$ and $X(\mbox{\amm})$ with $N(\mbox{H$_2$})$ (calculated in \S 4.4 assuming $T_d = T_K$) in Oph B, C and F. We additionally analyse Oph B1 and B2 separately to examine any potential differences between the two. In Figure \ref{fig:trends}a, we find that $T_K$ values in Oph C are nearly universally lower than those in the other filaments, and show a tendency to decrease with increasing H$_2$ column density. As described previously, the other Cores are warmer but also do not show a significant trend with $N(\mbox{H$_2$})$. We show in Figure \ref{fig:trends}b that Oph B1 and B2 are both consistent with having a constant, mildly supersonic ratio of non-thermal to thermal line widths over all $N(\mbox{H$_2$})$. Oph C line widths are generally dominated by thermal motions, and $\sigma_{NT}\,/\,c_s$ decreases significantly at  $N(\mbox{H$_2$}) \gtrsim 4 \times 10^{22}$\,cm$^{-2}$. It is interesting to note that there are few data points at these column densities in the other Cores, and no observed decrease in $\sigma_{NT}\,/\,c_s$. Figure \ref{fig:trends}c and d show that in all Cores $N(\mbox{\amm})$ tends to increase with $N(\mbox{H$_2$})$, following the general trend that the \amm\, emission follows the continuum emission in the Cores. The differences between \amm\, and continuum clumps are due to small scale differences in maxima. In addition, the fractional \amm\, abundance, $X(\mbox{\amm})$, tends to {\it decrease} with increasing $N(\mbox{H$_2$})$, although we note that Oph B1 and F have relatively few data points. Figure \ref{fig:trends}c also illustrates the high $N(\mbox{\amm})$ values in Oph B1 relative to $N(\mbox{H$_2$})$ compared with values in Oph B2, C and F. 

If \amm\, is depleting at high densities, as the variation of $X(\mbox{\amm})$ with $N(\mbox{H$_2$})$ suggests, \amm\, may not trace well the $T_K$ in the densest and likely coldest regions. Hence, our assumption of $T_d = T_K$ may not be valid towards the highest columns and we may be underestimating $T_d$ and $N(\mbox{H$_2$})$ by factors of $\sim 2$. For example, \citet{stamatellos07} predicted that $T_d$ in the centers of the Oph Cores could be as low as $\sim 7$\,K. Furthermore, $X(\mbox{\amm})$ ($\propto N(\mbox{H$_2$})$, see Figure \ref{fig:trends}) may be overestimated by similar factors. High resolution multiwavelength continuum observations and models are needed to obtain independent assessments of $T_d$ throughout the Oph Cores. 

\subsubsection{$\sigma_{NT}\,/\,c_s$ in Oph B}

\begin{figure}
\plotone{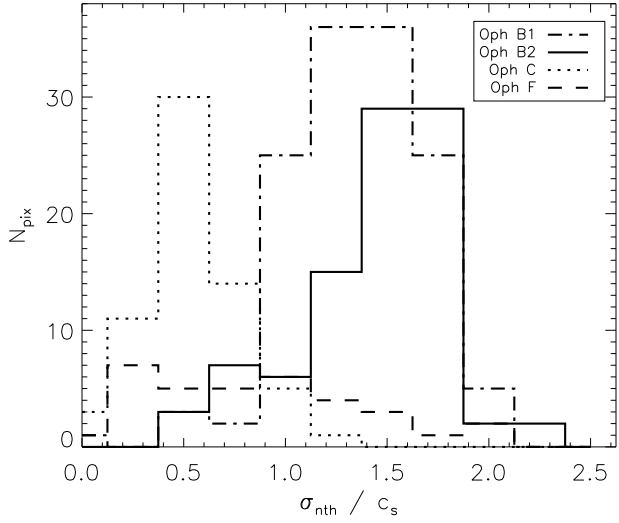} 
\caption{Histogram $\sigma_{NT}\,/\,c_s$ in Oph B1, B2, C and F. Values plotted are those in 15\arcsec\, pixels. Oph C is characterized by significantly lower non-thermal to thermal line width ratios than the other Cores. When analysed separately, Oph B1 and B2 are very similar in $\sigma_{NT}\,/\,c_s$.}
\label{fig:pixel_hist}
\end{figure}

\citet{jijina99} compiled numerous observations of \amm\, in dense gas and found that while most starless clumps are characterized by largely thermal motions, a large fraction of clumps in clusters have $\sigma_{NT} > \sigma_T$, and that the identification of an \amm\, clump as part of a cluster has a larger impact on the observed line widths than association with a protostar. These findings are in agreement with our results in Oph B. The line widths in Oph B, with a mean $\Delta v = 0.89$\,\kms\, and $\sigma_{NT}\,/\,c_s = 1.5$, are significantly wider and more dominated by non-thermal motions than those found in isolated cores. While B2 is associated with at least two embedded protostars and B1 appears starless, when studied separately (shown in Figure \ref{fig:pixel_hist}), both Cores have similar $\sigma_{NT}\,/\,c_s$ distributions. 

If the non-thermal component is caused by turbulent motions in the gas, then it is interesting to consider the turbulence source.  In the following, we consider the source of wide lines in this region as being due to ``primordial'' (i.e., undamped) turbulence, protostar-driven turbulence, bulk motions or biased sampling. 

Firstly, Oph B may have non-thermal motions throughout all the Core that are inherited from the surrounding cloud and that have not yet been damped. Since Oph B is associated with a few embedded protostars, however, it is likely that parts of the Core, at the very least, have been at high density for over a free-fall time $t_{ff}$ ($\sim 3 \times 10^5$\,years for $n(\mbox{H}_2) \approx 10^4$\,\cc, $\sim n_{cr}$ for the \amm\, (1,1) and (2,2) transitions). Since the dissipation timescale for turbulence is $\sim t_{ff}$ \citep{maclow04}, it is unlikely that the non-thermal motions from the parent cloud have been retained, if the embedded protostars are indeed physically connected to the Core. Determining the relative velocities of the YSOs compared with the Core $v_{LSR}$ would help to address this question. 

Secondly, the embedded protostars in Oph B may be adding turbulence to the core through energy input associated with mass loss. One outflow has been found in CO emission in the region associated with one of the two protostars  near the peak \amm\, integrated intensity in B2 (IRS45/GY273 and IRS47/GY279, labelled in Figure \ref{fig:b-compare}), with a blue lobe towards the west and a red lobe towards the south-east \citep{kamazaki03}. We do not, however, see localized regions of wide line widths associated with these or any protostars in Oph B. In fact, protostars are associated with $\Delta v$ {\it minima}. Additionally, wide line widths are found in B1, where there are no embedded protostars and no known outflows. This suggests that the large non-thermal motions across the core are not driven by embedded YSOs. 

Thirdly, the wide line widths seen in the \amm\, emission may be indicative of global infall in the Oph B Core. Given the mean \amm\, line width in Oph B2, we calculate a virial mass $M_{vir} \sim 8$\,M$_\odot$, assuming a density distribution which varies as $\rho \propto r^{-2}$. Given mass estimates of the Cores by \citet{motte98} (which are uncertain to factors $\sim 2$) and accounting for the different Ophiuchus distance used by the authors (160\,pc compared with our preferred value of 120\,pc), we find $M/M_{vir} \sim 5$ in Oph B2. Since the mean \amm\, (1,1) opacity $\tau = 1.6$ over Oph B, the average individual hyperfine component is optically thin, and we would not expect to see the asymmetrically blue line profiles found in collapsing cores in optically thick line tracers. \citet{andre07} find these spectroscopic signatures of infall motions in B2, with a clear blue infall profile towards our B2-A10 \amm\, clump, and profiles suggestive of infall towards B2-A7 and other continuum objects in its eastern half. No evidence of infall motions in the tracers used were found towards central B2, but the lines they used (CS, H$_2$CO and HCO$^+$) may suffer from depletion at the high densities and low temperatures found at this location, masking any infall signature. Oph B1, however, is also characterized by wide \amm\, line widths and has $M/M_{vir} \sim 1$. Additionally, spectroscopic infall signatures were found in Oph C towards our C-A3 \amm\, clump by \citeauthor{andre07}, where we find extremely narrow \amm\, line widths (similarly, B2-A7 contains the narrowest lines observed in B2). It does not appear, then, that infall motions are the sole contribution to the wide lines we find in Oph B. 

Lastly, rather than acting as a tracer of the densest gas in Oph B, the \amm\, emission may be dominated in this high density environment by emission from the more turbulent outer envelope of the Core. Modelling of the expected emission given a constant abundance of \amm\, would be necessary to determine accurately whether depletion is a factor in Oph B (note that we see evidence for decreasing abundance of \amm\, with $N(\mbox{H$_2$})$; see Figure \ref{fig:trends}d). Alternatively, observations of species which are excited at higher densities than $10^4$\,\cc, such as N$_2$H$^+$ and deuterated species such as N$_2$D$^+$ and H$_2$D$^+$, may better probe the dense clump gas. Moderate resolution ($\sim 26$\arcsec) resolution observations of N$_2$H$^+$ (1-0) \citep{andre07} find narrower line emission in small scale \dia\, condensations (which we call `clumps') in Oph B1 and the filament connecting B1 and B2 ($\sigma_{NT} = 0.15\pm0.04$\,\kms, or $\sigma_{NT}\,/\,c_s \sim 0.65$ for $T_K = 15$\,K), but non-thermal line widths in Oph B2 condensations remain transsonic ($\sigma_{NT}\,/\,c_s \sim 1.1 < 2$ on average). In an upcoming paper, we present H$_2$D$^+$ $1_{10} - 1_{11}$ observations in B2, showing that non-thermal line widths of gas at densities of $\sim 10^6$\,\cc\, (approximately the critical density of H$_2$D$^+$, depending on the collisional cross section used) in the core are also transsonic, $\sigma_{NT} \sim 0.25$\,\kms, or $\sigma_{NT}\,/\,c_s \sim 1.4$ at 10\,K (or $\sim 1.1$ at 15\,K; Friesen {\it et al.} 2009, in preparation). Regardless of whether or not the line widths seen in \amm\, emission are tracing the highest density gas, the mean non-thermal motions of gas in B2 are greater than typically found in isolated cores.



\subsubsection{$\sigma_{NT}\,/\,c_s$ in Oph C}

Motions in the Oph C Core differ substantially from those found in Oph B. Oph C is likely starless, like Oph B1. In contrast to B1 (and also Oph B2 and F), Oph C is dominated by nearly thermal motions ($\langle \sigma_{NT}\,/\,c_s \rangle = 0.6$), shown in Figure \ref{fig:pixel_hist}. Oph C thus appears less affected by the clustered environment, and is kinematically more alike isolated clumps. \amm\, clumps in the Pipe nebula and Perseus molecular cloud have similarly narrow line emission, for example, with the $\sigma_{NT}\,/\,c_s \sim 1-2$ in the Pipe \citep{rathborne08} and intrinsic line widths $\sigma_v$ typically less than 0.2\,\kms\, in Perseus \citep{rosolowsky08a}. 

In addition, we see in Oph C a progressive decrease in the magnitude of non-thermal motions as we look at molecular lines that trace increasingly high densities. Non-thermal line widths $\sigma_{NT} = 0.20 - 0.28$\,\kms\, were determined from DCO$^+$ (1-0) observations (albeit with a large 1$^\prime$.5 beam) of Oph C \citep{loren90}. At their peak emission locations in C, we find $\sigma_{NT} = 0.16$\,\kms\, and 0.17\,\kms\, for C$_2$S and HC$_5$N, respectively, from our GBT-only data. These locations coincide with the bulk of the \amm\, emission in C, where we find the smallest $\sigma_{NT}$ values (consistent at our velocity resolution with nearly or purely thermal motions), and where recent \dia (1-0) observations by \citet{andre07} also find $\sigma_{NT} \sim 0.13$\,\kms\,  (their C-MM3 - C-MM6). This trend indicates that turbulent motions have decreased at higher densities. 

 \subsubsection{Oph B3}

Oph B3 is an unusual object within Oph B. It was originally detected in DCO$^+$ emission \citep{loren90}. It is not readily apparent in dust continuum maps of Oph B, and consequently is only identified as a separate clump by \citet{stanke06}. This lack of prominence suggests that interesting potential sites of star formation may be overlooked by purely continuum surveys. (In a similar result, \citet{difran04} found a region of extremely narrow \dia (1-0) line width in Oph A associated with extended thermal continuum emission but not correlated with a continuum clump.) The relatively large difference in line of sight velocity between B3 and the Oph B mean (e.g., greater than that between Oph B and F) suggests that B3 may not be physically associated with the rest of Oph B. 

B3 is characterized in \amm\, emission by moderately bright, extremely narrow lines which are nearly thermal in width. The lack of continuum emission implies a low H$_2$ column density, leading to an \amm\, fractional abundance lower limit similar to the $X(\mbox{\amm})$ values found for clumps in B1 and C. Interestingly, no C$_2$S or HC$_5$N emission was detected in B3, suggesting that B3 may be in a later evolutionary state than B1 and C. Alternatively, if B3 is physically distinct from the rest of Oph B, its initial chemistry may have differed. 

The kinetic temperature associated with the B3-A1 \amm\, clump is 13.9\,K. This is lower than the average $T_K$ in Oph B but not particularly cold compared with other \amm\, clumps in Oph B. The uncertainty in this value is large due to the low signal-to-noise ratio of the \amm\, (2,2) line in the combined data, but we find a slightly lower value, $T_K = 12.0 \pm 0.3$\,K, from the single-dish data alone. This temperature difference is not large enough to account for the lack of submillimeter emission at the B3 peak if the column density of material was similar to that found in B1 or B2. 

Based on the flux observed by \citet{stanke06} at the location of B3 (their MMS-108) and the gas temperature derived from our observations, we find a total clump mass for Oph B3 of $M \sim 0.4\,M_\odot$, which is within a factor of $\sim 2$ (i.e., within uncertainties) of the virial mass calculated using the observed \amm\, line width and core radius from Table \ref{tab:peakdat} ($M \sim 0.2\,M_\odot$). Accordingly, if it collapses, Oph B3 may form a very low mass star or brown dwarf, depending on how much material passes into a compact protostar. Future sensitive large format millimetre array detectors, such as SCUBA-2, should easily detect objects like B3 at greater than 3-$\sigma$ levels of confidence with relatively short integration times. 

\subsection{Implications for Clustered Star Formation}

Comparisons of the mean values of many parameters found in each of the Cores (Tables \ref{tab:linefit} and \ref{tab:tketc}) show significant differences between Oph B and C. In particular, Oph C is characterized by significantly narrower line widths and lower kinetic temperatures. This distinction is further illustrated in Figures \ref{fig:trends} and  \ref{fig:pixel_hist}.  An object like Oph C (and perhaps Oph B3) resembles isolated clumps in the close correspondence between the \amm\, line map and the dust continuum map, and in the evidence that the peak of the intensity map coincides with local minima in both $T_K$ and $\Delta v$. Such a core would fit well the idea that clustered star formation is just a spatially concentrated version of isolated star formation, with smaller, denser star forming clumps packed closer together than in isolated regions. The Oph B Core shows a remarkably different behavior. Oph B contains a larger number of \amm\, clumps than Oph C or F, and there is much less correspondence between the locations of the \amm\, and continuum clumps. The \amm\, clumps show modest contrast with interclump gas in their intensity, and little contrast in their velocity or their line width. The greater amount of fragmentation of Oph B than found in C or F may be related to the relatively higher levels of turbulence in the Core, which are significantly greater than typically found in isolated regions. Thus, dense \amm\, gas in Oph B does not resemble the dense gas in regions of isolated star formation, and this raises the issue whether this presents a different `initial condition' for clustered star formation.
 
 
\section{Summary}

We have presented combined single-dish and interferometer \amm\, (1,1) and (2,2) observations of the B, C and F Cores in the clustered star-forming Ophiuchus molecular cloud. We additionally present single-dish C$_2$S ($2_1 - 1_0$) and HC$_5$N (9-8) observations of the Cores. Our main results can be summarized thus: 

1. While the large-scale features of submillimeter continuum emission and \amm\, (1,1) integrated intensity appear similar, on 15\arcsec\, scales we find significant discrepancies between the dense gas tracers in Oph B, but good correspondence in Oph C and F. We find poor correspondence in Oph B between continuum clumps and \amm\, clumps identified with 3D {\sc clumpfind}, with only five of fifteen \amm\, clumps located within 30\arcsec\, (3600\,AU) of a dust clump. This is in contrast with previous findings of extremely high spatial correlation between the two gas tracers in isolated, low-mass starless clumps.  

2. We find $v_{LSR}$ varies little across any of the Oph Cores, and additionally varies by only $\sim 1.5$\,\kms\, between them.  

3. Overall, the observed \amm\, line widths in the Cores are generally large, and often slightly supersonic. We also find regions of localized narrow line emission ($\Delta v \lesssim 0.4$\,\kms), some of which are associated with \amm\, clumps. The larger line widths in Oph B ($\langle \Delta v \rangle = 0.8$\,\kms) agree with previous findings for clumps in clustered regions. Line widths in Oph C, however, decrease to nearly thermal values which are more representative of typical isolated clumps. 

4. The derived kinetic temperatures of the gas are remarkably constant across Oph B. Kinetic gas temperatures in both B and F are warmer ($\langle T_K \rangle = 15$\,K) than generally found in isolated star forming clumps, but are consistent with temperatures determined for cores in clustered environments. The center of Oph C shows a minimum  $T_K \sim 9$\,K, similar to previous results in isolated clumps. 

5. We find no significant difference in $T_K$ between \amm\, clumps, submillimeter clumps ($\langle T_K \rangle = 14$\,K) and protostars ($\langle T_K \rangle = 15.5$\,K). Most other physical parameters have similarly insignificant variations between the objects. We do find that protostars are associated with significantly {\it smaller} line widths ($\langle \Delta v \rangle = 0.4$\,\kms) with approximately equal contribution from thermal and non-thermal motions. 

6. We have determined \amm\, abundance values towards the Cores, and find they agree with previous estimates of $X(\mbox{\amm})$ in cold, dense environments. Single-dish observations of C$_2$S and HC$_5$N resulted in only a few detections, and derived column densities ($\sim 10^{12} - 10^{13}$\,cm$^{-2}$) similar to those found in other molecular clouds. 

7. It is unlikely that the wide line widths observed in Oph B are due to pervasive turbulent motions inherited from the parent cloud if the already formed embedded protostars are physically connected with the core. We find no evidence of influence by the protostars on the gas motions, i.e., through local increases in line widths or gas temperatures. The \amm\, abundance in Oph B2 appears to decrease with increasing gas density. We therefore suggest that the \amm\, emission is biased, i.e., due to depletion at high inner densities, and is therefore not tracing the densest gas in the Core. This may explain the differences between the locations of \amm\, and continuum clumps. Consequently, the gas temperatures may be lower in the center of Oph B than given by the \amm\, line ratios. 

\acknowledgments

We thank the anonymous referee for thoughtful comments which improved this paper. We thank H. Kirk for providing SCUBA maps of the regions observed and D. Johnstone for useful discussions. We thank the observatory staff at the ATCA, VLA and GBT for their assistance in making our observations successful. In particular, we thank J. Ott for aid in setting up the ATCA observations and J. Lockman for providing scripts to grid OTF data from the GBT. The National Radio Astronomy Observatory is a facility of the National Science Foundation operated under cooperative agreement by Associated Universities, Inc. The Australia Telescope is funded by the Commonwealth of Australia for operation as a National Facility managed by CSIRO. The James Clerk Maxwell Telescope is operated by the Joint Astronomy Centre on behalf of the Particle Physics and Astronomy Research Council of the United Kingdom, the Netherlands Association for Scientific Research, and the National Research Council of Canada. RKF acknowledges financial support from the University of Victoria and the National Research Council Canada Graduate Student Scholarship Supplement Program. We also acknowledge the support of the National Science and Engineering Research Council of Canada. 

{\it Facilities:} \facility{ATCA}, \facility{GBT}, \facility{VLA}

\appendix

\section{Determining Physical Parameters from HFS Line Fitting Results}
\label{ap_line}

\subsection{Kinetic Temperature}

Since the metastable states across $K$-ladders are coupled only by collisions, and if we neglect the upper ($J \neq K$) non-metastable states, then the populations of molecules in the metastable states can be described by the Boltzmann equation.

\begin{equation}
\frac{n(J',K')}{n(J,K)} = \frac{g(J',K')}{g(J,K)} \exp\biggl(-\frac{\Delta E(J',K';J,K)}{T_{rot}(J',K';J,K)}\biggr)
\end{equation}

where $T_{rot}(J',K';J,K)$ is the rotational temperature relating the populations in the $(J,K)$ and $(J',K')$ states, and $\Delta E(J',K';J,K)$ is the energy difference between the two states. For $(J',K') = (2,2)$ and $(J,K)=(1,1)$, $\Delta E(2,2;1,1)/k = -41.5$\,K. The state statistical weights, $g(J,K)$ and $g(J',K')$, are equal to 3 and 5 respectively for the \amm\, (1,1) and (2,2) states \citep{hotownes}. Assuming the molecular cloud is homogenous such that $n(J',K')\,/\,n(J,K) = N(J',K')\,/\,N(J,K)$, we can solve for $T_{rot}(2,2;1,1)$ using

\begin{eqnarray}
\label{eqn:trot}
T_{rot}^{21} = T_{rot}(2,2;1,1) &=& -41.5 \biggl[ \ln\biggl(-\frac{0.283}{\tau(1,1,m)}\frac{\Delta \nu (2,2)}{\Delta \nu (1,1)}\\\nonumber
&\times& \ln\biggl[1 - \frac{\Delta T_A (2,2,m)}{\Delta T_A (1,1,m)}\biggl(1 - \exp(-\tau(1,1,m))\biggr)\biggr]\biggr)\biggr]^{-1}
\end{eqnarray}

\noindent where $\Delta T_A (1,1,m)$ and $\Delta T_A (2,2,m)$ are the antenna temperatures of the main component of the (1,1) and (2,2) transitions, respectively, $\tau (1,1,m)$ is the opacity of the (1,1) line summed over the main component only, and we assume that the line widths $\Delta \nu (1,1) = \Delta \nu (2,2)$. We then calculate the gas kinetic temperature $T_K$ from the rotational temperature following the updated result given by \citet{tafalla02} in their Appendix B:

\begin{equation}
\label{eqn:tk}
T_K = \frac{T_{rot}^{21}}{1 - \frac{T_{rot}^{21}}{42} \ln[1 + 1.1\,\mbox{exp}(-16/T_{rot}^{21})]}
\end{equation}

\subsection{Column Density}

We calculate the column density of the (1,1) transition following \citet{rosolowsky08a}:

\begin{equation}
N(1,1) = \frac{8\pi\nu_0^2}{c^2}\frac{g1}{g2}\frac{1}{A(1,1)}\biggl[1 - \exp{\frac{h\nu_0}{kT_{ex}}}\biggr]^{-1} \int{\tau(\nu)d\nu} ,
\label{eqn:column}
\end{equation}

\noindent where $g1 = g2$ are the statistical weights for the upper and lower states of the (1,1) inversion transition. The Einstein A coefficient $A(1,1) = 1.68 \times 10^{-7}$\,s$^{-1}$ \citep{pickett98}, and $\int{\tau(\nu)d\nu} = \sqrt{2\pi} \sigma_v / (c \nu_0) \tau_{tot}$. The total \amm\, column density $N(\mbox{\amm})$ can then be determined by calculating the value of the partition function $Z$ of the metastable states:

\begin{equation}
Z = \sum_J (2J+1) S(J) \exp{\frac{-h[B J (J+1)+(C-B)J^2]}{kT_k}} ,
\end{equation}
 
\noindent where the total \amm\, column is then $N(1,1) \times Z/Z(1,1)$. The values of the rotational constants B and C are 298117\,MHz and 186726\,MHz, respectively \citep{pickett98}. The function $S(J)$ accounts for the extra statistical weight of the ortho- over para-\amm\, states, with $Z = 2$ for $J=3,6,9,...$ and $Z = 1$ for all other $J$.  

\section{Consequences of 0.3\,\kms\, Velocity Resolution}
\label{ap_res}

To fit the hyperfine components of the \amm\, (1,1) inversion emission and obtain robust measurements of the opacity and excitation temperature of the line, we required an observational bandwidth wide enough to contain the main and at least two satellite components of the line. To meet this requirement, we were able to obtain velocity resolution across the band of only 0.3\,\kms\, due to the present correlator capabilities at the VLA. Here, we investigate the effect of the relatively low spectral line resolution on our determined line widths and kinetic temperatures by creating model spectra, including random noise with an rms value equal to that in our observed spectra. The spectra are then convolved by a finite spectral resolution and resampled. We then fit the resulting spectra with our custom HFS routine. 

\amm\, (1,1) spectra were modelled given an excitation temperature $T_{ex}$ and opacity $\tau$ typical of the regions observed, while the line width $\Delta v$ was varied. The line intensity at each velocity is then given by Equations \ref{eqn:tastar} and \ref{eqn:tau}. The amplitudes of the 21 components of the \amm\, (2,2) line were then determined in a similar manner, with the relative intensities of each component given by \citet{kuko67}. The relative intensity of the (2,2) line at a given $T_K$ was determined using Equation \ref{eqn:trot}, by first calculating the associated rotational temperature $T_{rot}$ (reversing Equation \ref{eqn:tk}). 

While the observational resolution does significantly impact some of the returned parameters, we show that the derived kinetic temperatures in this work are robust. 

\subsection{Line widths}

\begin{figure}
\plotone{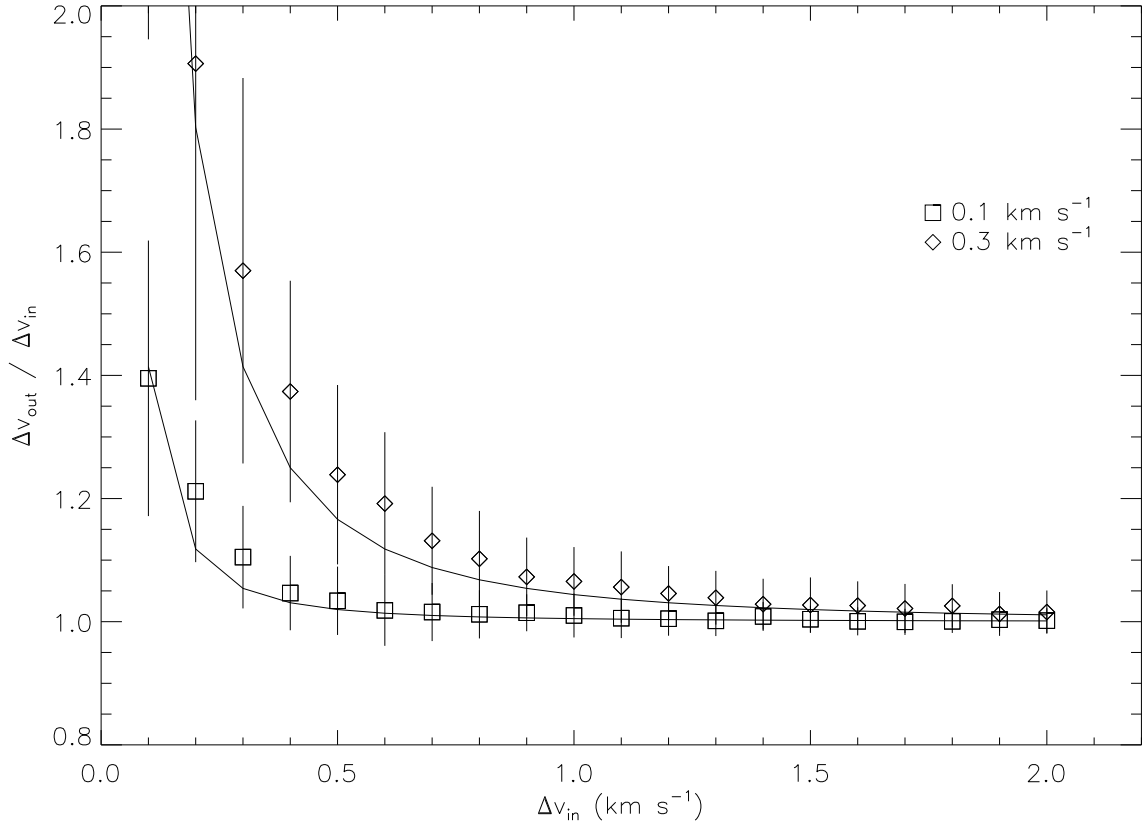} 
\caption{Input versus recovered line widths determined by creating a model spectrum convolved to 0.1\,\kms\, and 0.3\,\kms\, velocity resolution and subsequently fitting with the \amm\, HFS routine. Solid lines show the expected observed line width $\Delta v_{obs} = \sqrt{\Delta v_{line} + \Delta v_{res}}$ for 0.1\,\kms\, and 0.3\,\kms.}
\label{fig:fwhm_test}
\end{figure}

Figure \ref{fig:fwhm_test} plots the returned $\Delta v$ against that of the model spectrum for 0.1\,\kms\, and 0.3\,\kms\, resolution for a range of line widths. For both resolutions, the returned line width is greater than the actual for small values, with the 0.3\,\kms\, resolution showing a significantly greater effect. The trends, however, follow that expected for addition in quadrature of the true line width and the resolution, i.e. $\Delta v_{obs} = \sqrt{\Delta v_{line} + \Delta v_{res}}$ with some additional offset at small values and larger scatter due to poor resolution. For the mean values found in this study, the corrected line widths consequently reflect the true line widths, but at small widths the corrected values still overestimate the true value. For this reason, we cannot accurately state the true widths of lines where returned (uncorrected) values $\Delta v \lesssim 0.35$\,\kms, but do show that these uncorrected line widths are consistent with purely or nearly thermal values. 

\subsection{Opacity}

\begin{figure}
\plotone{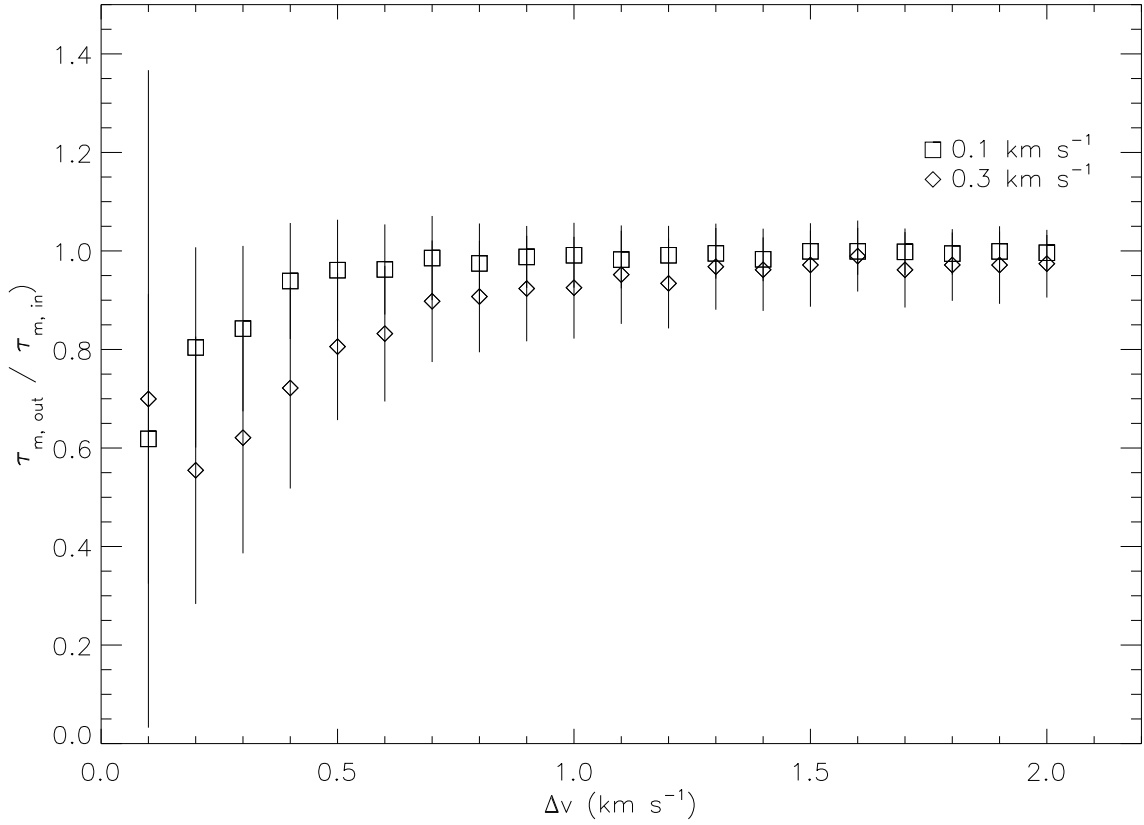} 
\caption{Recovered versus input \amm\, (1,1) opacities as a function of line width $\Delta v$ for 0.1\,\kms\, and 0.3\.\kms\, velocity resolution. At small line widths, the returned opacities are significantly smaller than the true values. }
\label{fig:tau_test}
\end{figure}

Figure \ref{fig:tau_test} plots the returned opacity $\tau$ against that of the model spectrum for 0.1\,\kms\, and 0.3\,\kms\, resolution over a range of line widths. The relatively low spectral resolution of our observations has a significant impact on the returned opacity, with our fitting routine significantly underestimating the opacity by greater than 20\% for line widths $\lesssim 0.7$\,\kms. In addition, the scatter in the returned opacities is large at small line widths, and this affects the uncertainty in the calculated \amm\, column density, $N(\mbox{\amm})$, in \S 4.3. Since $N(\mbox{\amm})$ depends linearly on the opacity (see Equation \ref{eqn:column}), the uncertainty in the opacity dominates the uncertainty in the column density at small line widths. Since the column density also depends linearly on line width, however, and the returned line width at small values is greater than the true value by a similar relative amount, the returned column density is more accurate than the uncertainties suggest. 

\subsection{Kinetic Temperature}

\begin{figure}
\plotone{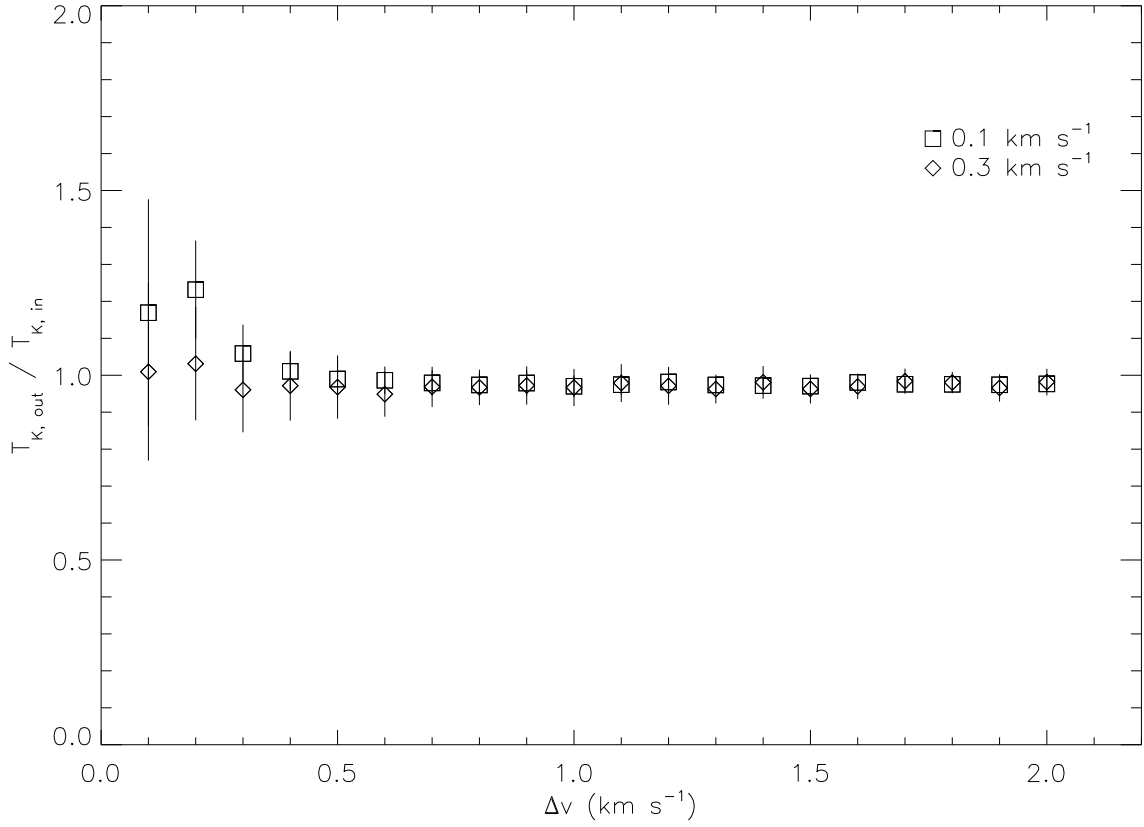} 
\caption{Recovered versus input kinetic temperatures $T_K$ as a function of \amm\, (1,1) line width $\Delta v$ for 0.1\,\kms\, and 0.3\,\kms\, velocity resolution. Even at small line widths, the returned kinetic temperature closely agrees with the input value, with little scatter.}
\label{fig:tk_test}
\end{figure}

The results from the above tests show that at small \amm\, line widths, the returned line widths and opacities can vary significantly from the true values. Regardless, the returned kinetic temperatures from our \amm\, line fitting are robust, as shown in Figure \ref{fig:tk_test} (for $T_K = 15$\,K). Even at small line widths ($\Delta v \lesssim 0.3$\,\kms), the kinetic temperatures derived from the fits are accurate within uncertainties of the true $T_K$, with relatively little scatter. At the lowest temperatures found in the Oph cores, $T_K = 10$\,K, the results are similar with slightly more scatter at the lowest line widths.  


\begin{deluxetable}{llll}
\tablecolumns{4}
\tablewidth{0pt}
\tablecaption{Rest frequencies of all observed spectral lines \label{tab:restfreqs}}
\tablehead{
\colhead{Molecule} & \colhead{Transition} & \colhead{Rest Frequency}  & \colhead{Source} \\
\colhead{} & \colhead{} & \colhead{GHz} & \colhead{} }
\startdata
C$_2$S         	& $(2_1 - 1_0)$            & 23.3440330 & \citet{pickett98}  \\
NH$_3$          	& $(1,1)$                     & 23.694495 & \citet{hotownes}  \\
NH$_3$          	& $(2,2)$                     & 23.722633 & \citet{hotownes}  \\
HC$_5$N	       	& $(9-8)$                    & 23.963888 & \citet{mhb79}  \\
\enddata
\end{deluxetable}

\begin{deluxetable}{lcc}
\tablecolumns{3}
\tablewidth{0pt}
\tablecaption{GBT Observation Details by Region \label{tab:gbtobsII}}
\tablehead{
\colhead{Core} & \colhead{Area Mapped}                     & \colhead{rms} \\
\colhead{}              & \colhead{{\it arcmin} $\times$ {\it arcmin} } & \colhead{K ($T_{MB}$)}}
\startdata
Oph B1 & $3 \times 3$ / $4 \times 5.5$ \tablenotemark{a} & 0.05 / 0.07 \tablenotemark{b}\\
Oph B2 & $5 \times 4$                             & 0.08 \\
Oph B3 & $3 \times 3$                             & 0.07 \\
Oph C   & $3 \times 3$ / $4 \times 4$   \tablenotemark{a}    & 0.05 / 0.07 \tablenotemark{b}\\
Oph F    & $3 \times 3$                             & 0.06 
\enddata
\tablenotetext{a}{First values indicate the original OTF map size. Second values are the final map sizes, increased due to observed extent of emission.}
\tablenotetext{b}{First rms value is calculated over the central region where all observations overlap. Second value is calculated over the entire map.}
\end{deluxetable}

\begin{deluxetable}{lccccc}
\tablecolumns{6}
\tablewidth{0pt}
\tablecaption{ATCA and VLA Observation Details by Region \label{tab:intobs}}
\tablehead{
\colhead{}              & \colhead{}   & \colhead{ATCA} & \colhead{} & \colhead{VLA} & \colhead{} \\
\colhead{Core} & \colhead{Overlap Extent} & \colhead{$N_{mos}$\tablenotemark{a}} &  \colhead{rms} 
										 & \colhead{$N_{mos}$\tablenotemark{a}} &  \colhead{rms}\\
\colhead{}              &  \colhead{{\it arcmin} $\times$ {\it arcmin}} & \colhead{}  & {mJy\,beam$^{-1}$}
										       & \colhead{}  & {mJy\,beam$^{-1}$}}
\startdata
Oph B1 & $2 \times 3$ & 3   &  20 & 7 \tablenotemark{b}  & 13 \\
Oph B2 & $5 \times 4$ & 10 &  20 & 16\tablenotemark{c} & 12\\
Oph C   & $4 \times 4$ & 7   &  30 & 7 & 10 \\
Oph F    & $3 \times 2$ & 3   & 30 & 5 & 15 \\
\enddata
\tablecomments{The spectral resolution of the observations is 0.1\,\kms\, (ATCA) and 0.3\,\kms\, (VLA).}
\tablenotetext{a}{Number of individual pointings in mosaic observations.}
\tablenotetext{b}{Pointings for B1 were aligned and spaced to provide continuous coverage between B1 and B2.}
\tablenotetext{c}{Additional pointings added to the B2 core to include B3 in the overlap region.}
\end{deluxetable}

\begin{landscape}
\centering
\begin{figure}
\plotone{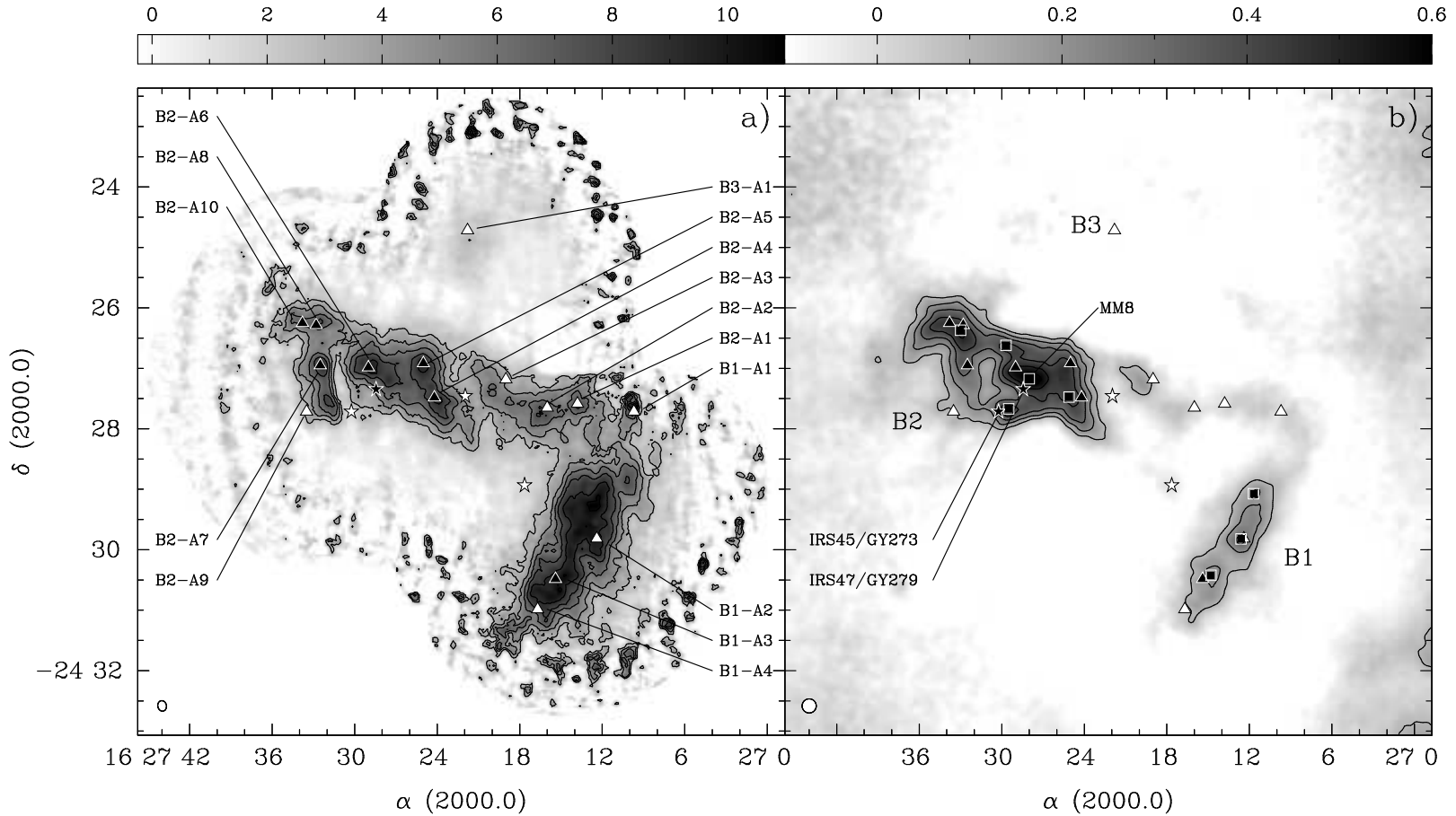} 
\caption{Observations of the Oph B Core (including B1, B2 and B3).  At lower left in each panel, ovals show the respective resolutions of the data. a) Integrated \amm\, (1,1) main component intensity obtained with the GBT, ATCA and VLA telescopes with a synthesized beam FWHM 10.5\arcsec\, $\times$ 8.5\arcsec. Emission was integrated over spectral channels with intensities $\geq \pm 2\,\sigma$ rms. The colour scale is in K\,\kms\, ($T_B$). Contours begin at 3\,K\,\kms\, and increase by 1.5\,K\,\kms. In both plots, stars denote locations of Class I protostars \citep{enoch08}, while triangles indicate the positions of \amm\, clumps as identified by {\sc clumpfind}. (b) Continuum emission at 850\,\micron\, in Oph B at 15\arcsec\, resolution as in \citet{jorgensen08}. The colour scale is in mJy\,beam$^{-1}$. Contours begin at 50\,mJy\,beam$^{-1}$ and increase by 100\,mJy\,beam$^{-1}$. Squares show locations of submillimeter clumps \citep{jorgensen08}. We also label the millimeter clump MM8 \citep{motte98} and protostars identified previous to \citet{enoch08}.}
\label{fig:b-compare}
\end{figure}
\end{landscape}

\begin{figure}
\plotone{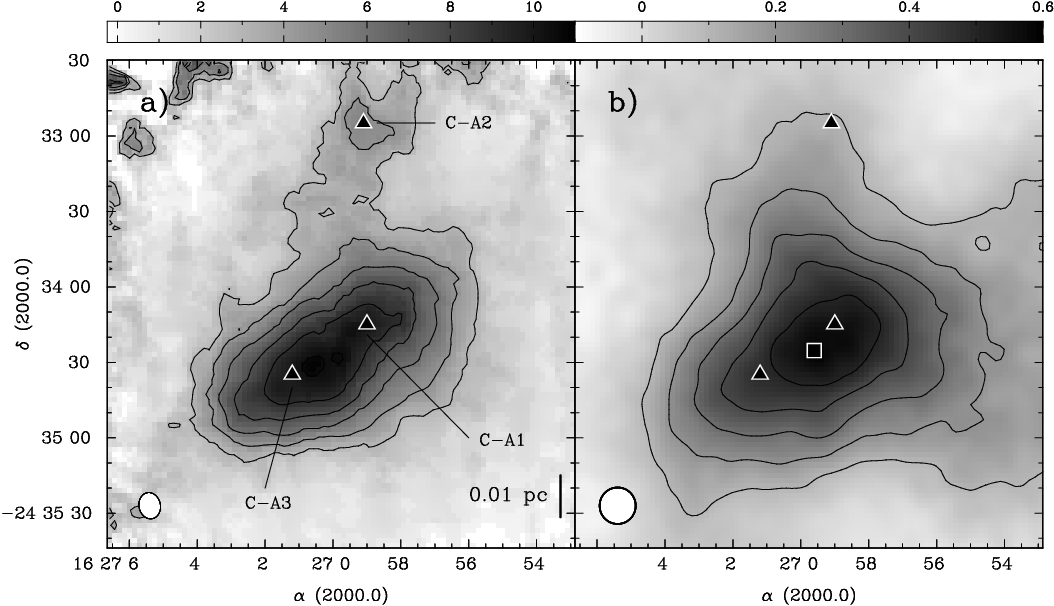} 
\caption{Observations of the Oph C Core. At lower left in each panel, ovals show the respective resolutions of the data. a) Integrated \amm\, (1,1) main component intensity obtained with the GBT, ATCA and VLA telescopes, summed over spectral channels with intensities $\geq 2\,\sigma$ rms. The colour scale is in units of K\,\kms\, ($T_B$). Contours begin at 3\,K\,\kms\, ($T_B$) and increase by 1.5\,K\,\kms. In both plots, triangles indicate the positions of \amm\, clumps as identified by {\sc clumpfind}. b) Continuum emission at 850\,\micron\, in Oph C at 15\arcsec\, resolution as in \citet{jorgensen08}. Contours and colour range as in Figure \ref{fig:b-compare}. Squares show locations of submillimeter clumps \citep{jorgensen08}.}
\label{fig:c-compare}
\end{figure}

\begin{figure}
\plotone{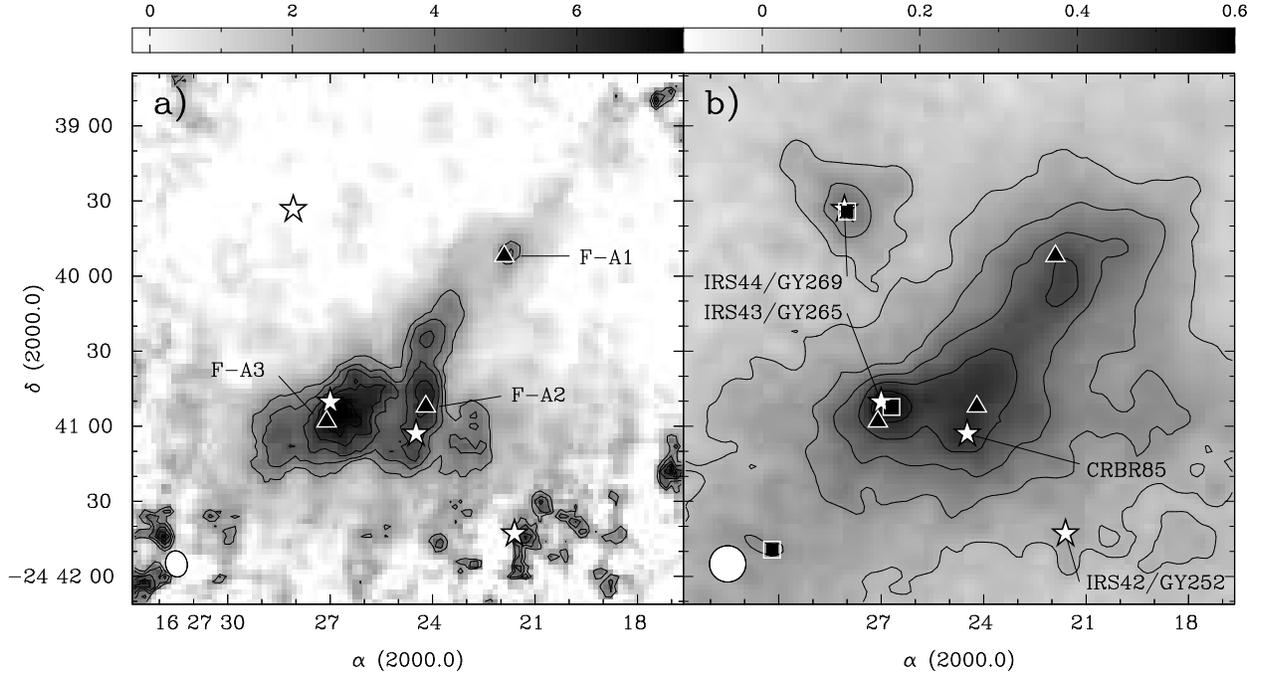} 
\caption{Observations of the Oph F Core. At lower left in each panel, ovals show the respective resolutions of the data. a) Integrated intensity \amm\, (1,1) main component intensity obtained with the GBT, ATCA and VLA telescopes, summed over spectral channels with intensities $\geq 2\,\sigma$ rms. The colour scale is in units of K\,\kms\, ($T_B$). Contours begin at 3\,K\,\kms\, ($T_B$) and increase by 1\,K\,\kms. In both plots, stars denote locations of Class I protostars \citep{enoch08}, while triangles indicate the positions of \amm\, clumps as identified by {\sc clumpfind}.  b) Continuum emission at 850\,\micron\, in Oph F at 15\arcsec\, resolution as in \citet{jorgensen08}. Contours and colour range as in Figure \ref{fig:b-compare}. Squares show locations of submillimeter clumps \citep{jorgensen08}. We also label protostars identified previous to \citet{enoch08}.}
\label{fig:f-compare}
\end{figure}

\begin{deluxetable}{ccccccccc}
\tablecolumns{9}
\tablewidth{0pt}
\tablecaption{GBT C$_2$S $2_1 - 1_0$ and HC$_5$N $9 - 8$ Peak Parameters \label{tab:ccsdat}}
\tablehead{
\colhead{ID} & 
\colhead{RA} & \colhead{Dec.} &
\colhead{FWHM} &
\colhead{$V_{lsr}$} & \colhead{$\Delta v$} &  
\colhead{$T_{MB}$} & 
\colhead{$N$} & \colhead{$X$} \\
\colhead{} & \colhead{J2000} & \colhead{J2000} & 
\colhead{AU} & 
\colhead{\kms} & \colhead{\kms} & 
\colhead{K} & \colhead{$10^{12}$ cm$^{-2}$} & \colhead{$10^{-10}$} }
\startdata
B1-C$_2$S	&  	16 27 13.1 & -24 30 50 &  5100 & 3.67(2) &  0.33(5)   &  0.36(3)  & 4.3(7)  & 3.1(7)\\
C-C$_2$S	& 	16 26 59.8 & -24 34 13 &  9200 & 3.95(1) &  0.33(2)   &  1.12(4)  & 10.6(8)  & 1.5(1)\\
C-HC$_5$N	&	16 26 58.8 & -24 34 30 &  5800 & 3.94(1) &  0.35(2)   &  0.68(3)  & 4.0(3) & 0.46(4)\\
\enddata
\end{deluxetable}

\begin{deluxetable}{lcccc}
\tablecolumns{5}
\tablewidth{0pt}
\tablecaption{\amm\, (1,1) clumpfind peaks and parameters \label{tab:peakdat}}
\tablehead{
\colhead{ID} & 
\colhead{RA} & \colhead{Dec.} &
\colhead{FWHM$_x \times$ FWHM$_y$\tablenotemark{a} } &
\colhead{Peak} \\
\colhead{} & \colhead{J2000} & \colhead{J2000} &
\colhead{(AU)} & \colhead{(K)}}
\startdata
B1-A1  & 16  27  9.7 & -24  27 43.0   & $ 3900 \times 6700 $ & 4.90 \\
B1-A2  & 16  27 12.4 & -24  29 49.0  & $ 6800 \times 7800 $ & 5.10 \\
B1-A3  & 16  27 15.4 & -24  30 29.1  & $ 5300 \times 5000 $ & 5.60 \\
B1-A4  & 16  27 16.7 & -24  30 59.1  & $ 3700 \times 2900 $ & 4.90 \\
B2-A1  & 16  27 13.8 & -24  27 35.0  & $ 2300 \times 2500 $ & 5.20 \\
B2-A2  & 16  27 16.0 & -24  27 39.0  & $ 5600 \times 6100 $ & 5.10 \\
B2-A3  & 16  27 19.0 & -24  27 11.0  & $ 4700 \times 6700 $ & 3.90 \\
B2-A4  & 16  27 24.2 & -24  27 29.0  & $ 6400 \times 6000 $ & 5.10 \\
B2-A5  & 16  27 25.0 & -24  26 55.0  & $ 5300 \times 4700 $ & 5.00 \\
B2-A6  & 16  27 29.0 & -24  26 59.0  & $ 3800 \times 6700 $ & 5.70 \\
B2-A7  & 16  27 32.5 & -24  26 57.0  & $ 3800 \times 5200 $ & 6.70 \\
B2-A8  & 16  27 32.8 & -24  26 17.0  & $ 3600 \times 3600 $ & 4.00 \\
B2-A9  & 16  27 33.5 & -24  27 43.0  & $ 5000 \times 3600 $ & 3.60 \\
B2-A10 & 16  27 33.8 & -24  26 14.9 & $ 2700 \times 3700 $ & 4.00 \\
B3-A1  & 16  27 21.8 & -24  24 42.9  & $ 3900 \times 6100 $ & 3.60 \\
C-A1   & 16  26 59.0 & -24  34 14.8   & $ 5000 \times 5600 $ & 5.40 \\
C-A2   & 16  26 59.1 & -24  32 54.8   & $ 4500 \times 4400 $ & 4.30 \\
C-A3   & 16  27  1.2 & -24  34 34.8    & $ 6000 \times 5400 $ & 5.50 \\
F-A1   & 16  27 21.9 & -24  39 52.0   & $ 3600 \times 3200 $ & 4.50 \\
F-A2   & 16  27 24.2 & -24  40 52.1   & $ 4400 \times 5900 $ & 4.90 \\
F-A3   & 16  27 27.1 & -24  40 58.1   & $ 2400 \times 5000 $ & 5.50 \\
\enddata
\tablenotetext{a}{FWHM values calculated by {\sc clumpfind} have not been deconvolved with the synthesized beam of the combined data.}
\end{deluxetable}

\begin{deluxetable}{llllll}
\tablecolumns{6}
\tablewidth{0pt}
\tablecaption{\amm\, (1,1) Line Characteristics in Combined Data \label{tab:linefit}}
\tablehead{
\colhead{Core} & \colhead{Value} & \colhead{Mean} & \colhead{RMS} & \colhead{Min} & \colhead{Max}}
\startdata
Oph B  &  $v_{lsr}$ (\kms)  	&  3.96  &  0.24  &  3.15  & 4.57 \\
              &  $\Delta v$ (\kms) 	&  0.83  &  0.21  &  0.08  & 1.37 \\
              &  $\tau$       			&  1.6    &  0.8    &  0.5     & 4.7    \\
              &  $T_{ex}$  (K) 		&  9.9    &  2.3    &  5.4     & 19.5 \\ \hline
Oph C  &  $v_{lsr}$  (\kms) 	&  4.01  &  0.07  &  3.81  & 4.10 \\
              &  $\Delta v$ (\kms) 	&  0.37  &  0.13  &  0.11  &  0.70 \\
              &  $\tau$             		&  4.7    &  2.6    &  0.7   &  10.8     \\
              &  $T_{ex}$  (K)      	&  6.9    &  1.2    &  5.1   &  11.6  \\ \hline
Oph F  &  $v_{lsr}$   (\kms)	&  4.34  &  0.12  &  4.24  & 4.82 \\
              &  $\Delta v$ (\kms)	&  0.53  &  0.30  &  0.10  &  1.13 \\
              &  $\tau$       			&  1.9    &  0.8    &  0.7   &  3.6     \\
              &  $T_{ex}$  (K)		&  9.5    &  2.5    &  5.0   & 14.3 \\
\enddata
\end{deluxetable}

\begin{deluxetable}{llcccc}
\tablecolumns{6}
\tablewidth{0pt}
\tablecaption{Physical Properties of Filaments Derived From Fitted Parameters \label{tab:tketc}}
\tablehead{
\colhead{Core} & \colhead{Value} & \colhead{Mean} & \colhead{RMS} & \colhead{Min} & \colhead{Max}}
\startdata
Oph B        & $T_k$ (K) 				& 15.1 & 1.8   & 12.3 & 22.9 \\
		 & $\sigma_{NT}$ (\kms)		& 0.35 & 0.09 & 0.04 & 0.57 \\
		 & $\sigma_{NT}\,/\,c_s$		& 1.5	   & 0.4    & 0.5   & 2.5  \\
		 & $N(\mbox{\amm})$ ($\times \,10^{13}$\,cm$^{-2}$) 	& 22 & 8.8 & 1.3 & 48 \\
		 & $n(\mbox{H}_2$) ($\times \,10^4$\,\cc) 				& 5.9 & 10 & 0.8 & 77 \\
		 & $X(\mbox{\amm}$) ($\times \, 10^{-9}$)				& 14 & 9.1 & 2.4 & 41 \\ \hline
Oph C	 &  $T_k$ (K) 				& 12.0 & 1.6   & 9.4   & 19.8 \\
		 & $\sigma_{NT}$ (\kms)		& 0.14 & 0.06 & 0.0 & 0.29  \\
		 & $\sigma_{NT}\,/\,c_s$		& 0.6   & 0.2    & 0.0 & 1.2  \\
		 & $N(\mbox{\amm})$ ($\times \,10^{13}$\,cm$^{-2}$) 	& 24 & 11 & 7.1 & 55 \\
		 & $n(\mbox{H}_2$) ($\times \,10^4$\,\cc) 				& 2.0 & 1.0 & 0.8 & 4.2  \\
		 & $X(\mbox{\amm}$) ($\times \, 10^{-9}$)				& 8.2 & 4.4 & 2.8 & 28 \\ \hline
Oph F	 & $T_k$ (K) 				& 15.5 & 2.5   & 12.8 & 22.9 \\
		 & $\sigma_{NT}$ (\kms)		& 0.21 & 0.13 & 0.05 & 0.47 \\
		 & $\sigma_{NT}\,/\,c_s$		& 0.9    & 0.6   & 0.2   & 2.2  \\
		 & $N(\mbox{\amm})$ ($\times \,10^{13}$\,cm$^{-2}$) 	& 14 & 5.9 & 3.5 & 25 \\
		 & $n(\mbox{H}_2$) ($\times \,10^4$\,\cc) 				& 3.6 & 2.4 & 0.9 & 10.6 \\
		 & $X(\mbox{\amm}$) ($\times \, 10^{-9}$)				& 5.4 & 1.8 & 2.2 & 10 \\
\enddata
\end{deluxetable}

\begin{landscape}
\begin{deluxetable}{lcccccccccc}
\tablecolumns{11}
\tablewidth{0pt}
\tablecaption{Derived Parameters at \amm\, (1,1) Peak Locations \label{tab:peak_dat}}
\tablehead{
\colhead{ID} &
\colhead{$v_{lsr}$} & 
\colhead{$\Delta v$} & 
\colhead{$\tau$} & 
\colhead{$T_k$} & 
\colhead{$T_{ex}$} & 
\colhead{$\sigma_{NT}$} &
\colhead{$N(\mbox{H}_2)$} &
\colhead{$N(\mbox{NH}_3)_{tot}$} &
\colhead{X(\amm)} &
\colhead{$n(\mbox{H}_2$)} \\
\colhead{} &
\colhead{km\,s$^{-1}$} &
\colhead{km\,s$^{-1}$} & \colhead{} &
\colhead{K} & 
\colhead{K} & 
\colhead{km\,s$^{-1}$} &
\colhead{$10^{21}$\,cm$^{-2}$} &
\colhead{$10^{13}$\,cm$^{-2}$} &
\colhead{$10^{-9}$} &
\colhead{$10^4$\,\cc}}
\startdata            
B1-A1  &  4.01(3) & 0.85(7) & 1.6(3) & 16.3(1.9) &   9.4(1) & 0.35(3) &  \nodata &  22(4)  & \nodata &  2.8(1)  \\
B1-A2  &  3.84(2) & 0.92(5) & 1.5(1) & 16.4(1.3) &  12.1(1) & 0.38(2) & 14(3) &  29(4) &  17(8) &  6.5(2) \\
B1-A3  &  3.98(1) & 0.62(3) & 3.4(2) & 13.7(1.0) &   9.7(1) & 0.25(2) & 18(4) &  41(3) &  23(9) &  4.2(1) \\
B1-A4  &  3.83(2) & 0.54(3) & 4.2(3) & 12.4(0.9) &   8.4(1) & 0.22(2) & 12(2) &  42(3) &  35(12) &  3.2(1) \\ 

B2-A1  &  4.18(2) & 0.57(5) & 1.6(3) & 13.3(1.2) &  11.6(1) & 0.23(2) & \nodata &  22(4) & \nodata & 13(1) \\
B2-A2  &  4.13(2) & 0.86(5) & 0.3(*) & 14.1(*) &  19.5(1) & 0.35(*) & \nodata &  18(*) & \nodata &  0.0(*) \\
B2-A3 &  4.10(3) & 0.80(7) & 1.3(3) & 14.2(1.2) &  10.2(1) & 0.33(3) &  15(2) &  20(4) &  14(6) &  4.7(1) \\
B2-A4  &  4.00(2) & 1.00(6) & 1.4(2) & 13.5(1.0) &  11.6(1) & 0.42(3) & 32(6) &  32(5) &  9.7(4) & 11.7(8) \\
B2-A5  &  3.91(2) & 0.79(5) & 2.4(2) & 13.2(1.0) &  9.7(1) & 0.32(2) & 45(9) &  38(4) &  8.6(3) & 4.8(2) \\
B2-A6  &  4.32(2) & 0.66(4) & 1.3(2) & 14.5(1.1) &  14.1(1) & 0.27(2) & 33(7) &  24(4) &  6.6(3) & 72(26) \\
B2-A7  &  4.57(1) & 0.33(2) & 2.8(2) & 13.9(1.1) &  11.9(1) & 0.12(2) & 35(7) &  22(2) &  6.5(2) & 11.5(8) \\
B2-A8  &  4.35(3) & 0.95(8) & 0.5(*) & 14.6(*) &  16.9(1) & 0.39(*) & 43(7) &  16(*) &   3.7(*) &  0.0(*) \\
B2-A9 &  \nodata & \nodata & \nodata & \nodata & \nodata & \nodata & \nodata & \nodata & \nodata & \nodata \\
B2-A10  &  4.22(2) & 0.71(6) & 1.6(3) & 14.3(1.2) & 10.3(1) & 0.29(3) & 35(7) &  23(4) &  6.1(3) &  4.8(2) \\ 

B3-A1 &  3.15(2) & 0.08(8) & 0.7(*) & 13.9(*)    &  13.9(1) & 0.0(*) & \nodata & 1.3(*) & \nodata & 12.4(*)  \\

C-A1  &  4.08(1) & 0.16(1) & 9.3(6) & 10.4(0.7) &   8.0(2) & 0.0(*) & 84(14) & 33(2) &  4.0(2) &  4.2(2) \\  
C-A2  &  3.97(1) & 0.36(4) & 2.7(6) & 12.0(1.2) &   8.8(2) & 0.13(1) & 14(3) & 20(4) &  14(9) &  4.1(2) \\  
C-A3  &  3.98(2) & 0.29(2) & 9.3(5) & 10.7(0.8) &   8.0(2) & 0.10(3) & 68(4) &  60(5) &  8.8(2) &  3.9(2) \\ 

F-A1  &  4.82(4) & 0.8(1)  & 1.5(7) & 15.3(2.0) &   7.4(2) & 0.32(5) & 31(6) &  16(7) &  5.1(4)  &  1.6(1) \\
F-A2  &  4.31(2) & 0.58(5) & 1.8(4) & 15.6(1.6) &   9.9(2) & 0.23(2) & 35(7) &  19(5) &  5.5(3)  &  3.4(1) \\ 
F-A3  &  4.31(2) & 0.43(3) & 2.6(4) & 16.8(1.8) &   9.8(2) & 0.16(2) & 24(5) &  19(3) & 7.8(3)  &  3.0(1) \\ 
\enddata 
\tablecomments{Uncertainties are given in brackets beside values and show the uncertainty in the last digit of the value, with the exception of the $T_{K}$ uncertainties which are given in K units. A (*) indicates that the uncertainty in the value at this position were large, even though in most cases the values themselves are reasonable. A good HFS fit was not found at the peak position of \amm\, clump B2-A9. }
\end{deluxetable}      
\end{landscape}                 

\bibliographystyle{apj}
\bibliography{biblio}

\end{document}